 \documentclass[preprint2]{aastex}
 
\newcommand{\msun}{{M$_\sun$}}
\newcommand{\kms}{{km s$^{-1}$}}
\newcommand{\myr}{Myr}

\newcommand{\eff}{{\rm{eff}}}
\newcommand{\snn}{{Starburst99}} 
\newcommand{\vrotcero}{$v_{\rm{rot}} = 0$}
\newcommand{\vrotthree}{$v_{\rm{rot}} = 300$}

\shorttitle{Massive Stellar Populations with Rotation}
\shortauthors{Gerardo A. V\'azquez et al.}

\begin{document}


\title{Models for Massive Stellar Populations with Rotation}


\author{Gerardo A. V\'azquez}
\affil{Physics \& Astronomy Department, Johns Hopkins University, 3400 N Charles St.,
    Baltimore, MD 21218}
\email{vazquez@pha.jhu.edu}


\author{Claus Leitherer}
\affil{Space Telescope Science Institute, 3700 San Martin Drive Baltimore, MD 21218}
\email{leitherer@stsci.edu}

\and

\author{Daniel Schaerer\altaffilmark{1}, Georges Meynet and Andre Maeder}
\affil{Observatoire de Gen\`eve, 51, Ch. des Maillettes, CH-1290 Sauverny, Switzerland}
\email{Daniel.Schaerer@obs.unige.ch, Georges.Meynet@obs.unige.ch, Andre.Maeder@obs.unige.ch}


\altaffiltext{1}{Laboratoire d'Astrophysique (UMR 5572), Observatoire Midi-Pyr\'en\'ees, 14 Avenue E. Belin, F-31400 Toulouse, France}

\begin{abstract}

We present and discuss evolutionary synthesis models for massive stellar populations generated with the Starburst99 code in combination with a new set of stellar evolution models accounting for rotation. The new stellar evolution models were compiled from several data releases of the Geneva group and cover heavy-element abundances ranging from twice solar to one fifth solar. The evolution models were computed for rotation velocities on the zero-age main-sequence of 0 and 300~{\kms} and with the latest revision of stellar mass-loss rates. Since the mass coverage is incomplete, in particular at non-solar chemical composition, our parameter study is still preliminary and must be viewed as exploratory. Stellar population properties computed with Starburst99 and the new evolution models show some marked differences in comparison with models obtained using earlier tracks. Since individual stars now tend to be more luminous and bluer when on the blue side of the Hertzsprung-Russell diagram, the populations mirror this trend. For instance, increases by factors of two or more are found for the light-to-mass ratios at ultraviolet to near-infrared wavelengths, as well as for the output of hydrogen ionizing photons. If these results are confirmed once the evolution models have matured, recalibrations of certain star-formation and initial mass function indicators will be required.

\end{abstract}





\section{Introduction}

Massive stars are among the most important cosmic stellar yardsticks in stellar populations at low and high redshift. Yet, in many regards they still pose serious challenges in evolutionary synthesis modeling. For instance, evolutionary connections with and between Wolf-Rayet (WR) stars, the significance of the nitrogen (WN) and carbon (WC) sequence among the WR stars, the ratio of blue to red supergiants, stellar surface abundances, or stellar mass loss are all poorly understood \citep{mas03}.

The Geneva and Padova groups perfected the state-of-the-art of stellar evolution modeling in the uppermost Hertzsprung-Russell diagram (HRD) with the release of complete set of stellar evolutionary tracks \citep{schll92,Bre93,Sch93a,sch93b,Ch93,bertelli94,fag94a,fag94b,mey94}. These sets have become widely used in population synthesis modeling \citep[e.g.,][]{fr97,S99,Ce02, schulz02,bch03,robert03,vazleith05}.

Previous modeling of massive star evolution paid particular attention to the size of the convective core and stellar mass loss, which were thought (together with standard nuclear processing) to be the most relevant ingredients in massive star evolution. This view has changed quite dramatically in recent years. It is become clear that stellar rotation plays a dominant -- and in some cases even {\em the} dominant -- role in the evolution of stars with masses above $\sim$10~{\msun} \citep{mame00a}. As a result, the previous generation of stellar evolution models, which do not take into account rotation, may be subject to revision. This in turn will affect all population synthesis modeling relying on these tracks.

Recently, a new set of tracks has been released by the Geneva group \citep[and references therein]{mema05}, which addresses stellar rotation. This set of tracks should be considered ``exploratory'', i.e., it is still incomplete and not meant to fully replace the currently available full set of tracks without rotation. These new tracks, however, are invaluable in exploring the fundamental effects caused by stellar rotation and gauging the
impact on stellar population modeling. Consequently we have implemented this new set of tracks into the latest version 5.1 of the evolutionary synthesis code {\snn} \citep{vazleith05}. The resulting new version of Starburst99 is not publicly available, as the limited number of evolutionary tracks does not yet allow the computation of models by the astronomical community that could be compared to a wide range of observations.

The present paper is organized as follows. In Section~2 we summarize the main astrophysical ingredients in the new stellar evolution models with rotation and highlight differences with respect to the previous non-rotating grid. We also discuss details of the implementation in Starburst99 in this section. A parameter study of the most important stellar population properties affected by rotation is performed in Section~3. In Section~4 we compare the model predictions obtained with Starburst99 to observations. Finally, the conclusions are in Section~5.

\section{Stellar tracks with rotation}

In this section we will describe the most important characteristics of the new set of Geneva tracks and highlight differences with respect to previous tracks. The discussion will be relatively brief and limited to aspects that are directly relevant to the implementation of the tracks into Starburst99.

\subsection{Physics of rotation}

The effects of rotation can be classified in three categories: 1) the hydrostatic effects, 2) the rotationally induced instabilities and their impacts on the transport of the chemical species and of the angular momentum, 3) the effects of rotation on mass loss. The hydrostatic effects account for the fact that a rotating star is deformed. The stellar structure equations are modified with respect to the non-rotating case. The method used here is the one proposed by \citet{kipp70}, adapted to the case of shellular rotation as explained in \citet{mema97}. The two main instabilities triggered by rotation are the meridional circulation and the shear instabilities. The effects of these two instabilities on the transport of the chemical species and the angular momentum are modeled as proposed by \citet{zahn92}. This theory describes the interactions of these two instabilities in a consistent way. It is based on one sole a priori hypothesis, namely that the star is in a state of shellular rotation (angular velocity $\Omega$ constant on isobaric surfaces). This hypothesis is quite reasonable from a physical point of view, since on an isobaric surface, there is no force counteracting the horizontal, {\em i.e.} isobaric, turbulence. The effects of rotation on mass loss have been included as prescribed by \citet{mame00a}. Note that anisotropic stellar winds were included only for the models at $Z =0.02$. The rotational velocities considered in these models are too modest for this effect to be important at other chemical compositions. More details on the physics of the models can be found in \citep{mema03,mema05}.

\subsection{Physical ingredients of the new models}

Extensive coverage of the physics of rotation in the Geneva tracks is provided in a series of papers published by the Geneva group \citep[e.g.,][and references to prior papers therein]{hirschi05, mema05}. The physical effects of rotation are discussed in detail in this series of
papers and will not be repeated here. The new stellar evolution models used in the present population synthesis modeling are those of \citep{mema03,mema05}. The rotating models account for the transport of the chemical species and of the angular momentum driven by meridional circulation and shear instabilities.

The models were computed with an initial equatorial velocity on the zero-age main-sequence (ZAMS) of 300 km~s$^{-1}$. Different processes during the main sequence phase (e.g., the expansion of outer layers, transport of angular momentum by stellar winds, etc.) decrease this velocity, so that the time-averaged velocity on the main sequence for different chemical compositions is 200~--~250~km~s$^{-1}$ \citep[see tables in][]{mema03,mema05}. 
Recently, \citet{dufton06} have measured stellar rotation velocities of Galactic B-type stars in NGC~3293 and NGC~4755 whose estimated ages are 10 and 15 Myr, respectively. Dufton et al. obtained a distribution of velocities that peaks at 250 {\kms} after accounting for a random distribution of the inclination for stars with masses between 3 and 12 {\msun}\footnote{On average, $v = 4/\pi v \sin i$}. This result is similar to that obtained by \citet{huang05}. On the other hand, \citet{mokiem06} obtained mean velocities of 150~--~180 {\kms} for O and B-type stars in the SMC with initial masses in the range of 15 to 30 {\msun}. In view of these recent results, we see that our averaged velocities might be overestimated for the SMC stars, while they seem to be in good agreement for Galactic stars.

Of course, for a proper comparisons with observations, one should compute models for the whole range of observed velocities and then convolve the results with the distribution of the initial velocities. We think that the results of the 300~km~s$^{-1}$ models will sufficiently well represent the averaged evolution so that meaningful comparisons can already be performed on the basis of these simple models.

The initial compositions were calculated as in \citet{mema01}, where $Y=Y_p + \Delta{Y}/\Delta{Z} \times Z$ with $Y_p=0.23$ and $\Delta{Y}/\Delta{Z}=2.5$. It follows that $X=0.757$, 0.744, 0.640 and $Y=0.239$, 0.248, 0.320 for heavy-element abundances of $Z=0.004$, 0.008, 0.040, respectively. In the case of solar chemical composition, the values are $X=0.705$ and $Y=0.275$. The heavy-element ratios were assumed to remain
constant with $Z$ and are the same as those used to calculate the opacities for solar composition. The nuclear reaction rates were taken from the NACRE data base \citep{angulo99}. A more detailed description of opacities and nuclear reactions can be also found in \citet{vazleith05}. Moderate overshooting is included in all the new models with and without rotation \citep{mema03}.

For the main sequence, the mass-loss rates are from \citet[2001]{vink00}, who take into account the abrupt change of the ionization and opacity of the stellar wind at certain temperatures, as explained by \citet{mame00a}. The empirical formula of \citet{dejager88} was used for the domain not covered by \citet{vink01}. For the WR phase, the mass-loss rates of \citet{nuglam00} were used. We note that our adopted main-sequence mass-loss rates were derived from theoretical wind models. \citet{puls06} found good agreement between these rates and those observed in O stars in the Galaxy and the Magellanic Clouds. The agreement is less satisfacory, however, for B supergiants whose observed rates are lower than predicted. These rates are empirical, as opposed to the theoretical main-sequence rates. The Nugis \& Lamers rates, which account for the clumping effects in the winds, are smaller by a factor of 2~--~3 than the rates used in the previous non-rotating ``enhanced mass-loss rate'' stellar grids. The correction factor for enhancing the theoretical mass loss due to
rotation was taken from Maeder \& Meynet. The dependence on chemical composition during non-WR phases was expressed as $\dot{M} \propto \dot{M}_\odot \times Z^{0.5}$ \citep{kudpul00}. WR mass-loss rates were assumed to be $Z$-independent in the default set of tracks. However, an additional set of tracks with $Z$-dependent mass-loss rates for the WR phase was computed for models with $Z=0.040$;
\citet{crowther02a} suggested the same dependence of the mass-loss rates on $Z$ for WR stars as for O stars ($\propto Z^{0.5}$). The mass-loss rates differ from those in the previous release of the Geneva tracks.

\subsection{Boundary conditions at the stellar surface}

The temperature distribution of a rotating star is described by the local radiative flux, which is proportional to the effective gravity $F \propto g_\eff$, where $g_\eff$ is the sum of the gravity and the centrifugal force \citep{vonzeipel}. Then the local effective temperature on the surface of the rotating star varies like $T_\eff(v) \propto g_\eff(v)^{1/4}$, where $v$ is
the rotational velocity. Due to this effect, the emergent luminosity, colors and spectrum will be different for different orientation angles {\em i} \citep{mape70}. In order to avoid such complications, we define an average stellar $T_{\eff}$ by $T_{\eff}^4 =  L/(\sigma S(\Omega))$, where $\sigma$ is Stefan's constant and $S(\Omega)$ the total actual stellar
surface deformed by rotation. This is the average value of the effective temperature which will be used for visualizing the evolutionary tracks in the theoretical HRDs and for evaluating the radiation outputs.

Since WR winds have  a non-negligible optical thickness, a correction is applied to new stellar models by adopting an effective radius $R_\eff$ at the optical thickness $\tau=2/3$. $R_\eff$ is related to the photospheric radius via the relation

\begin{equation}
R_\eff=R + {3{\kappa}\vert{\dot{M}}\vert  \over 8\pi v_{\infty}},
\end{equation}

where $\kappa$ is the opacity, $\vert\dot{M}\vert$ is the mass-loss rate, and $v_{\infty}$ is the wind terminal velocity. This procedure is similar to that followed in \citet{schll92} in the previous set of tracks. Subsequently, the effective temperature at $\tau=2/3$ is calculated with the relation $L=4\pi{R_\eff^2}{\sigma}{T_\eff^2}$.

The WR atmospheres are linked to the stellar models using the empirical relation of \citet{smith02} who defined the evolutionary temperatures as $0.6 T_\star + 0.4 T_{\eff}$, where $T_\star$ is the temperature of the WR core. This new definition is a compromise between the hotter interior and the cooler surface temperature and serves as the interface temperature used in the expanding model atmospheres. Smith et al. empirically determined this relation by comparing their model predictions to the observed temperature
distributions of Galactic WC and WN stars.

\subsection{The evolution of the rotation velocity}

In the interior of the star, the angular velocity varies due to the following effects: contraction/expansion of the stellar layers and transport of angular momentum by convection, meridional currents and shear instabilities. Without the transport processes, the change of $\Omega$ would only be governed by the
local conservation of the angular momentum. In that case, there is no coupling between the various stellar layers, each one keeping its initial angular momentum content.
In contrast, the transport processes impose some coupling between them.
During the main-sequence phase and in the outer part of the star, the transport mechanisms relocate angular momentum from the inner to the outer regions. This tends to slow down the contracting core and to accelerate the expanding outer envelope. When the stellar winds are weak, as for instance in metal-poor regions, the angular momentum can accumulate in the outer layers so that the surface velocity increases with time, approaching the critical limit. When the stellar winds are strong, the angular momentum of the outer layers is removed and the surface velocity decrease, evolving away from the critical limit. This illustrates the importance of the mass loss for shaping the evolution of the surface velocities of massive stars.

\subsection{Consequences for the evolutionary tracks}

The differences between the previous and the new evolutionary models result from a combination of effects. First and foremost, rotation introduces the majority of the changes. In addition, the opacities and overshooting are different. The former produce a shift in temperature towards lower effective temperatures. The overshooting is responsible for lower effective temperatures during the main-sequence, and combined with a core extension due to rotation it also affects the blue loops during the He-burning phase. Other major effects are:

\begin{itemize}

\item{Tracks with rotation become more luminous than those with no rotation. As a consequence, the derived mass of a star on the HRD will be lower than compared with non-rotating tracks.}
\item{The main-sequence width is enlarged due to rotation and overshooting for stars with initial masses below $\sim 30$~{\msun}. This effect results from a larger He-core produced during the H-burning phase.}
\item{For masses higher than 50 {\msun}, the opposite effect occurs. For instance, the reddest point reached by the 60 {\msun} non-rotating stellar model at the end of the main-sequence phase is $\log{T_{\eff}} \sim 4.15$, whereas for the rotating model, it is $\sim 4.6$. This is a consequence of the lower diffusion time scale ratio for chemical elements at higher initial mass. Massive stars will be more mixed than stars with lower masses at the
same evolutionary stage. A higher degree of homogeneity is reached, and this turns the star bluer. Combined with these effects, rotation and mass loss change the stellar surface composition and favor the formation of WR stars.}
\item{The mass-loss rates used in the present models are smaller than the high rates used in \citet{mey94}. Therefore, the final masses at the end of the He-burning phase are larger than those in \citet{schll92}.}
\item{ The main-sequence life times at $Z=0.020$ and 0.04 have increased by 15~--~25 and 16~--~27\% in comparison with the non-rotating models, respectively. This is mainly an effect of rotational mixing which continuously supplies the core with fresh nuclear fuel, i.e., hydrogen. The increase of the luminosity due to rotation cannot compensate for this effect, and the main-sequence life times are increased by rotation, at least in this metallicity range.}

\end{itemize}

 \subsection{Predicted properties of WR stars}

The WR phase is significant for massive stars, and stellar rotation can dramatically affect this phase. Our knowledge of this phase is insufficient to permit constraints on the properties of WR stars purely from first principles \citep{mema03,mema05}. Therefore some ad hoc assumptions must be made. For the present stellar models, the star enters into the WR phase when 1) $\log{T_{\eff}} \geq 4.0$ and 2) when the mass fraction of hydrogen at the surface is $X_{\rm{s}} < 0.4$. There is no significantly different outcome if the value for $X_{\rm{s}}$ varies between 0.3 or 0.4. In the first release of {\snn} \citep{S99}, the definition of a WR star was based on the work of \citet{mae91}, who used a minimum temperature limit for a WR star of $\log{T_{\eff}} \ge 4.4$. This limit is 0.4~--~0.5 dex higher than that assumed in the new tracks. Recent observations of WC/WN ratios in the Magellanic Clouds resulted in better agreement with a lower temperature limit \citep{mas03}. The WR definition via Teff is arguably rather adhoc and needs to be addressed. Therefore we have studied synthesis models with both temperature limits and with the old and new sets of tracks to gauge the importance of
this parameter for the WR properties.

\begin{figure}
\plotone{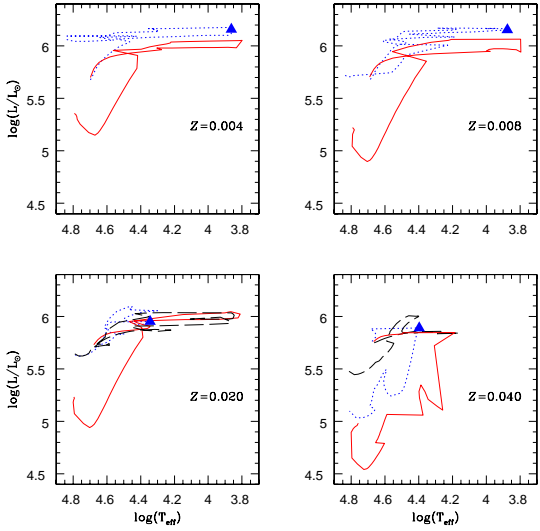}
\caption{Comparison of three 60 {\msun} tracks with different physics input from the Geneva group; solid: the original high mass loss, short-dashed:  {\vrotcero}, dotted: {\vrotthree}~{\kms}. The triangle indicates the beginning of the WR phase for the {\vrotthree}~{\kms} models.
\label{fig1}}
\end{figure}

We are also interested in the stellar mass at the entrance and during the WR phase. As a reminder, the most massive stars enter the WR phase while still in the H-burning phase. Two different mass limits for the WR phase have been defined by the Geneva group \citet{mema03,mema05}. The first one is the mass limit to enter to the WR phase on the main-sequence ($M_{\rm{OWR}}$); the second mass limit is the mass to enter the WR phase at any time during the evolution of the star ($M_{\rm{WR}}$). The first limit implies an evolutionary scenario avoiding the luminous blue variable (LBV) phase so that the phases covered are O-eWNL-eWNE-WC/WO.\footnote{The transition from the WN to the WC phase occurs when the mass fraction of carbon is higher than 10\% of the mass fraction of nitrogen. ``e'' denotes a spectral subclass defined by evolutionary criteria \citep{foell03}.} The second mass limit implies an evolutionary scenario first going through the O-LBV or
red supergiant (RSG) phases and then following the sequence eWNL-eWNE-WC/WO. The mass limits for different sets of tracks are shown in Table~\ref{tbl-1}. No rigorous comparison with observations has been made but the sparse data suggest a higher $M_{\rm{OWR}}$ closer to 75 {\msun} for Galactic stars \citep{crowther07}.
Fig.~\ref{fig1} shows evolutionary tracks for 60 {\msun} with an indication of the entry point into the WR phase for four different chemical compositions for the new models with {\vrotthree} {\kms}. The tracks for models with {\vrotcero} {\kms} and the previous tracks with high mass-loss rates are also included in Fig.~\ref{fig1}.

\subsection{Evolutionary phases and implementation into {\snn}}

The principal new set of tracks has an initial $v_{\rm{rot}}$ of 300 {\kms} with four different values of the chemical composition, as described in Section~2.2. A second set of tracks has $v_{\rm{rot}} = 0$~{\kms}. The non-rotating models were used to calibrate and compare the new models with rotation to the previous sets of tracks. These tracks were generated for $Z=0.02$ and 0.04. An additional set of tracks with $Z=0.04$ and metallicity dependent WR mass-loss rates is used in this paper as well.

The tracks were computed up to the end of the He-burning phase in all cases. Table~\ref{tbl-2} summarizes the covered mass range for the different metallicities. The tracks with solar chemical composition have the most complete coverage of parameter space. Tracks were released in the same way as the previous set of tracks by considering equivalent evolutionary time-step points along each track. The number of points, however, increases
by a factor of 7. Point 1 defines the ZAMS, point 100 is the end of the H-burning phase, point 201 is the beginning of the He-burning phase, and point 350 is the end of the He-burning phase.

The new set of tracks predict the same stellar parameters as in the original set of the Geneva tracks. New parameters like the ratio of the polar to the equatorial radii, equatorial velocity and angular velocity ($\Omega/\Omega_{breakup}$) are also included in the new release. Throughout the paper, we will use the following nomenclature: the ``old Geneva tracks'' are the previous tracks with high mass-loss rates used as default by {\snn}. The ``new Geneva tracks with or without rotation'' are the new tracks with $v_{\rm{rot}} = 300$~{\kms} and 0~{\kms}, respectively.

The method of implementation into {\snn} was similar to the one used by \citet{vazleith05}. However, the number of grid points of the new tracks is different. In order to be consistent with the existing, old set of tracks, we carefully selected 77 grid points to define key evolutionary phases and cover periods of rapid evolutionary changes. To achieve consistency, we found it necessary to resample the old Geneva tracks from the original 51 to 84 grid points. The old tracks extend up to the C-burning phase. We re-built the new tracks with 84 points as well, even though their evolution does not go beyond the He-burning phase. This leaves 7 slots available for a future implementation of the C-burning phase.

\section{Rotation versus non-rotation}

The purpose of this section is to compare integrated parameters from Starburst99 obtained with the old and new tracks. Constant star formation and instantaneous burst models were generated with an initial mass function (IMF) index of $\alpha = 2.35$ over the mass range of 1.0~--~100 {\msun} until 100 Myr. We used the fully line-blanketed atmospheres of \citet{smith02} for hot stars and ATLAS9 models adjusted by \citet[1998]{le97} for stars of spectral type B and later.

\subsection{Isochrones}

\begin{figure}
\plotone{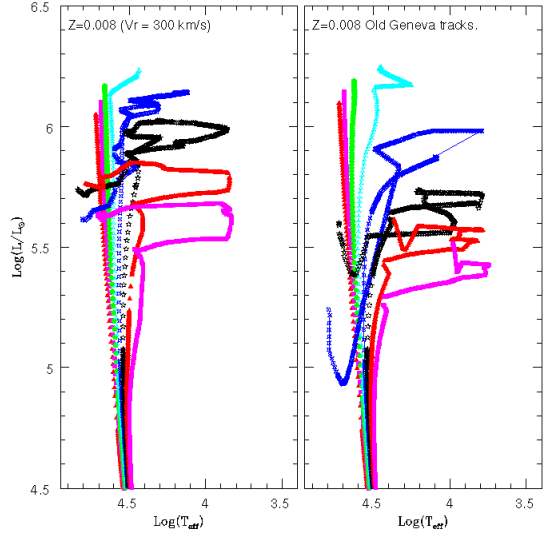}
\caption{Isochrones produced with Starburst99. $Z = 0.008$: ages of 0.01, 1, 2, 3, 4, 5, 6, and 7~Myr. $Z = 0.020$: ages of 0.01, 1, 2, 5, 8, 11, 14, 17, 20, 23, 26, and 29~Myr. The oldest isochrone in each figure corresponds to the lifetime of the least massive star included in the tracks with {\vrotthree}~{\kms}.
\label{fig2}}
\end{figure}

As a first comparison, we show in Fig.~\ref{fig2} isochrones for two different abundances ($Z = 0.008$ and 0.020) produced with the new tracks with rotation of {\vrotthree} {\kms} (left) and those produced with the old Geneva tracks (right). In each case the oldest isochrone corresponds to the age at which the lowest mass included in the new models with {\vrotthree}~{\kms} is present (see Table~\ref{tbl-2}).

Isochrones produced with the new tracks are more luminous than those obtained with the old Geneva tracks. Stars on the RSG branch with $Z = 0.008$  are bluer than those obtained with the old Geneva tracks. In the case of $Z = 0.02$, the new isochrones do not have the extended blue loop except for the model with 9 {\msun}. As this is the lowest mass model included in the new set of tracks, this isolated blue loop for just one mass will create a bump in the integrated light properties, as we will see below. \citet{mema03} discuss how the reduction of the blue loops in the new tracks with rotation results from the combined effect of overshooting and the core extension by rotation.

\subsection{Spectral energy distributions}

We have compared the stellar spectral energy distributions (SEDs) for young ages. Throughout this paper we follow the convention to plot the models with the different sets of tracks from the Geneva group with enhanced mass loss, the new models with {\vrotcero}, and the new models with {\vrotthree} {\kms} with solid, short-dashed and dotted lines, respectively. Young population models with $Z=0.020$ in the range of 1-15 Myr were divided and are plotted in two different figures. Fig.~\ref{fig3} shows models up to 6 Myr, and Fig.~\ref{fig4} shows the models with older ages. SEDs produced by the models with rotation show an energy increase around the neutral helium edge with respect to the models with old Geneva tracks, which can be up to $\sim 5$ dex. This is the result of a higher effective temperature for a given age in the new models. The flat sections at the shortest wavelengths of some spectra in Fig.~\ref{fig3} and Fig.~\ref{fig4} are numerical artifacts.

The difference between the old and new models in the range of $\sim 2.4 \le \log{\AA} \le 2.95$ at ages 3~--~7~Myr may not appear very striking with the scaling in Figures~\ref{fig3} and \ref{fig4}. Nevertheless, it amounts to about a factor of two for the number of ionizing photons, as we will discuss below. The onset and disappearance of the enhanced ionizing luminosity coincides with the presence of WR stars in the new models with rotation.

At even longer wavelengths above $\sim 10000$ {\AA} (Figure~\ref{fig5}), the difference in luminosity is $< 0.5$ dex for the same young ages covered in Figs.~\ref{fig3} and \ref{fig4}. When the age increases and RSGs begin to dominate the red and near-infrared part of the SED, the models with rotation become even more luminous as they predict a larger number of RSGs.

We restrict our discussion of the SEDs to the case of $Z = 0.02$ because only this set of evolutionary models is sufficiently complete to allow the computation of a meaningful SED.

\subsection{Massive-star inventory}

\subsubsection{The WR/O ratio}

The first test for studying the effects of rotation, mass loss, and temperature change is the ratio of the number of WR to O stars for an instantaneous burst. Fig.~\ref{fig6} shows this ratio for six different instantaneous burst models: old Geneva tracks and $\log{T_{\eff}} \ge 4.4$ (red solid line); old Geneva tracks and $\log{T_{\eff}} \ge 4.0$ (red long-dashed); new tracks of {\vrotthree} {\kms } and $\log{T_{\eff}} \ge 4.4$ (blue dotted); new tracks of {\vrotthree} {\kms } and $\log{T_{\eff}} \ge 4.0$ (blue dot-dashed); new tracks of {\vrotcero} {\kms} and $\log{T_{\eff}} \ge 4.0$ (black short-dashed); new tracks of {\vrotthree} {\kms}, $\log{T_{\eff}} \ge 4.0$ and different mass mass loss rate for WR phases at $Z = 0.040$ (dot long-dashed). As a word of caution, the new models with lower chemical composition than solar are incomplete at the lowest masses. While the WR evolution is well covered using the new tracks, the evolution of the least massive O stars is not.

The WR-rich phase after an instantaneous starburst covers a larger interval of ages when rotation is accounted for. This comes from the fact that rotational mixing aids an entry at an earlier evolutionary stage. In order to appreciate this point better, recall that mass loss by stellar winds is the key process allowing nuclearly processed layers to appear at the surface and thus make the star enter into the WR phase in non-rotating models. In the case of rotating models, another effect contributes in building the chemical characteristics of the stellar surface of WR stars: the mixing induced by rotation brings elements produced in the stellar core to the surface. Thus, for the present mass-loss rates, rotation favors an entrance into the WR regime at an evolutionary stage when the central hydrogen mass fraction is higher. As a consequence, the WR lifetimes will be longer. Also, stars with lower initial mass can be progenitors of WR stars. An immediate consequence are the peaks in the WR/O ratio between 6 and 9 Myr for $Z = 0.020$ and 0.040. These artifacts result when WR stars exist but most O stars have already evolved into B stars.

Furthermore, the WR phase lasts significantly longer in the new models so that the overall effect is a net increase of the WR/O ratio in an equilibrium population. This becomes obvious in Fig.~\ref{fig7}, where we have plotted the same ratio for constant star formation. The higher number of WR stars in the new models is significant.

Generally, there are more WR stars in the new stellar models with rotation and with the lower temperature limit for the WR phase. It is important to realize that most of this difference is caused by WR stars evolving from the lowest masses. These are the most populous species, and varying their numbers has the most dramatic effect.

\subsubsection{The WC/WN ratio}

The same cautionary remarks on incompleteness as in the previous subsection apply here. Fig.~\ref{fig8} shows the WC/WN ratio for instantaneous star formation. As in the case of the WR/O ratio, we plot models for constant star formation as well (Fig.~\ref{fig9}). The general trend in these figures is a decrease of WC/WN, in particular at low chemical composition. This effect is most dramatic for models with $Z=0.008$ and 0.004. As discussed by \citet{mas03}, the previous generation of Geneva tracks has been plagued by predicting a too high WC/WN ratio for metal-poor environments. The new models appear to be an improvement in this respect (cf.\ Fig.~\ref{fig27}).

The dot long-dashed line plotted for $Z=0.040$ in Figs.~\ref{fig8} and \ref{fig9} is a model for the WR phase with the new tracks of {\vrotthree} {\kms} using the lower temperature limit ($\log{T_{\eff}} \ge 4.0$) and a different prescription for the mass-loss rate during the WR phase. These tracks were specifically introduced to boost the predicted number of WC stars for higher chemical composition. However, contrary to expection, Figs.~\ref{fig8} and \ref{fig9} suggest that these models do not change significantly the result: the WC/WN ratio is slightly lower (dot-long dashed line) than in the corresponding new models without $Z$-dependent WR mass-loss rates (dotted line).

\subsubsection{The R/WR ratio}

Another often employed ratio is that of RSGs over WR stars (R/WR). Before exploring this ratio, it needs to be defined first. Our definition is based on the work of \citet{mas02} who considers as RSGs all stars with $(V - R)_0 > 0.6$, corresponding to stars of $\log{T_{\eff}} \leq 3.66$ (4600 K, or late G type) and at the same time having $M_{\rm{bol}}<-7.5$. This definition ensures that only very luminous red stars are counted and less luminous AGB stars are excluded. We have implemented this definition in {\snn} to calculate the predicted number of RSGs.

As in the previous sections we are taking into account different temperature limits for the WR phase. We do not discuss models with instantaneous star formation as R/WR is indeterminate in this case. The WR and RSG phase are not overlapping in the evolutionary models, and the ratio of the two quantities would be either zero or infinity. This degeneracy results from our extreme assumption of an idealized single stellar population. In the presence of a small age spread at star birth, there is a small likelihood of both RSGs and WR stars co-existing. Observations of clusters suggest few if any such examples. Fig.~\ref{fig10} shows the R/WR ratio for constant star formation. To understand this figure, we have to take into account the life times of massive stars, which enter when stellar populations in equilibrium are considered.

The panel for $Z=0.020$ suggests values of order unity for R/WR, with only a mild dependence on the adopted tracks or temperature definition. In the case of $Z = 0.040$, the spread between the different tracks becomes much larger. The difference can be up to $\sim 1$~dex, independently of the temperature limit for WR formation. The new tracks for this chemical composition predict a higher number WR stars (therefore a lower R/WR ratio) because the WR phase sets in very early close to the main sequence.

The panels for $Z=0.004$ and 0.008 should be interpreted with care because of the incompleteness of the tracks. The differences between the individual curves are caused by variations of the WR numbers only. Massive stars with $M \ge 40$ {\msun} will never reach the RSG phase in this part of the HRD. We will return to this point when we analyze the HRD and compare it with observations. On the other hand, those RSGs that would contribute to R/WR are mostly stars with initial masses $\le 25$ {\msun} but they are not included in the new models because of incomplete tracks (cf. Table~\ref{tbl-2}).

\subsubsection{The B/R ratio}

Next, we discuss the blue-to-red supergiant ratio (B/R). As in the previous section, we need to define this ratio first. We will again use the definition of \citet{mas02}, who defines blue supergiants as having $(B - V)_0 < 0.14$, corresponding to stars of $\log{T_{\eff}} \geq 3.9$ (8000 K, or type A9 I), and $M_{\rm{bol}}<-7.5$. As in the case of RSGs, we implemented
this definition into {\snn} to calculate the number of blue supergiants.

Fig.~\ref{fig11} shows the B/R ratio for constant star formation. The case of instantaneous star formation has been omitted for a similar reason as before: very luminous blue supergiants have no evolutionary counterpart in the uppermost HRD.

The trends in the individual panels are qualitatively similar to those for R/WR. The two panels for $Z=0.004$ and 0.008 are again to be taken with care. The apparent increase of B/R with decreasing $Z$ is an artifact of incomplete tracks. In contrast, the panels for $Z=0.020$ and 0.040 reflect the action of the new evolution  models and are suitable for comparison with observations.

\subsection{Ionizing continuum}

Figure~\ref{fig12} shows the number of ionizing photons below 912 {\AA}, 504~{\AA}, and 228~{\AA} for an instantaneous starburst. The photon output of the three sets of tracks is similar during epochs when WR stars are not present. In contrast, significant differences are observed in the presence of WR stars between 3 and 8 Myr. The models with rotation generate more ionizing photons than the original tracks. The effect is most pronounced for
the ionized He continuum, which was opaque with the old tracks but becomes more luminous by many orders of magnitude in the new models. Note, however, that the photon output in absolute terms is still very small for most of the time and may not be observable. The effect does become observable in the neutral He continuum, as well as in the neutral hydrogen continuum. The latter displays a factor of 2~--~3 increase over the old models. We emphasize the immediate consequences on, e.g., the calibration between the ionizing luminosity and the star-formation rate and its dependence on $Z$, as the WR/O ratio is strongly $Z$ dependent.

As a reminder, the treatment of $T_{\eff}$ at the interface with the extended atmospheres is the same as in \citet{schll92}, but $R_{\eff}$ at the location of the optical depth $\tau(R_{\eff}) = 2/3$ changes according to the new rotation physics. Furthermore, we have modified the temperature used in the expanding atmospheres as suggested by \citet{smith02}. Using a different recipe for the interface temperature would change the properties of the WR population. Although the recipe for the temperature is the same as in our previous Starburst99 model set, the WR parameters of the evolution models are quite different now. This explains why the ionizing photon output is very different, despite using the same WR model atmospheres.

\subsection{Panchromatic luminosities}

One of the most important consequences of rotation is that the new stellar models are more luminous at any wavelength. We compared the predicted luminosities from an instantaneous burst for different spectral bands. As before, the old Geneva tracks, new tracks of {\vrotthree} {\kms}, and {\vrotcero} {\kms} are represented by solid, dotted, and short-dashed lines, respectively. Figure~\ref{fig13} shows the bolometric
($M_{\rm{bol}}$), 1500 {\AA} ($L_{\rm{1500}}$), {\em B}-band ($M_B$), {\em V}-band ($M_V$), {\em J}-band ($M_J$), and {\em K}-band ($M_K$) luminosities for solar composition. The photometric system, the $L_{\rm{1500}}$ definition and the zero point for $M_{\rm{bol}}$ are as in \citet{vazleith05}.

We see from Fig.~\ref{fig13} that the luminosities are more affected by the new models with {\vrotthree} {\kms} than by those with {\vrotcero} {\kms}. The contribution to the bolometric and short-wavelength bands is mainly due to hot main-sequence and WR stars for young populations with ages $< 15$~{\myr}. In addition, all tracks have higher bolometric luminosity, thereby raising $L_{\rm{1500}}$. At longer wavelengths, the increased emission is due to the excursion of stars with masses $< 30$ {\msun} to the RSG phase for ages $> 20$ {\myr}. The RSG phase is significantly enhanced in the new models with rotation.

\subsection{Colors}

Fig.~\ref{fig14} displays the color evolution for an instantaneous burst at three different chemical compositions. The usual caveat as to incompleteness at low $Z$ applies. We decided to include results for metal-poor populations as they can provide important astrophysical insight when interpreted in context. The readers are urged to keep in mind the compilation of tracks in Table~\ref{tbl-2}.

The contribution from RSGs to the red bands is most obvious in ($V - R$) and ($V - K$) from models with {\vrotthree} {\kms} after 15 {\myr}. The effects on colors are delayed with respect of the old models. The new models with {\vrotcero} {\kms} follow the prediction of the old Geneva tracks except for younger ages $< 10$ {\myr}, when the new tracks evolve faster than the old ones.

In general, for all plots shown in Fig.~\ref{fig14}, the new models with {\vrotthree} {\kms} predict values of ($U - B$) bluer by up to $\sim 0.2$ magnitudes than the models using the old Geneva tracks. There is a higher contribution in the {\em U} band for populations $\le 7$ {\myr}. The ($B - V$) color is less affected, but still shows a noticeable effect. The difference between old and new models can be up to $\sim 0.1$ magnitudes during some epochs.

The major differences at solar chemical composition occur for ($V - R$) and ($V - K$). After 10~--~15 {\myr} these colors are redder by up to 1.5 magnitudes in the new models with rotation. As discussed before, this effect is due to the redward evolution to the RSG phase of stars with masses lower than 30 {\msun}. In the new models, the stars spend a longer fraction of their life times in the red part of the HRD, rather than evolving back to
higher temperatures on blue loops. The peak seen around ${\log}{\rm(Age [Myr])} \sim 1.6$ in Fig.~\ref{fig14}c is the effect of the blue loop for the model of 9 {\msun}. This loop is introduced by the interpolation in the mass range $7 < M/M_{\odot} < 12$ (see the discussion in Section 2).

\subsection{Mass loss and light-to-mass ratio}

In the absence of rotation, the prescription used to predict the mass-loss rate leads to lower rates on the main-sequence than in the old Geneva tracks. In combination with rotation, the mass-loss rates are enhanced by the time the stars enter the WR phase at the end of the main-sequence. Therefore, the total mass loss and all related parameters, such as the mass-to-light ratio, are different in comparison with the previous models. We have plotted in Fig.~\ref{fig15} the mass-loss rate of instantaneous burst models using the new tracks with rotation of {\vrotthree}, new tracks of {\vrotcero} {\kms} and models using the old Geneva tracks labeled as dotted, short-dashed, and solid lines, respectively. We have also included the set of tracks with $Z = 0.040$ with a mass-loss rate $\propto Z^{0.5}$ for the WR phase. This set is plotted as the dot-dashed line in the bottom right panel labeled as $Z = 0.040$.

Fig.~\ref{fig15} suggests somewhat higher mass-loss rates close to the ZAMS for models with rotation than for models with the old Geneva tracks. This effect is more pronounced at lower chemical composition. Subsequently, the new mass-loss rates are lower at the end of the main-sequence but will again surpass the rates of the original models in the WR phase. It is important to recall that the comparison in Fig.~\ref{fig15} is for the mass-loss rates integrated over the entire population at a given evolutionary phase, and not for the mass-loss rates at a given position in the HRD. The rates used for the new models with rotation during the WR phase would be lower than those used in the old Geneva tracks for a given position in the HRD. In the case of the new rotating models, since the stars have higher mass when they enter the WR phase, they have also higher luminosities and thus higher mass-loss rates.

The figure for $Z = 0.040$ includes an additional model (dot-dashed line). This model is based on tracks using a mass-loss rate $\propto Z^{0.5}$ for the WR phase. This scaling produces somewhat higher rates in the WR phase at super-solar chemical composition, an effect that can be clearly seen in the figure. While this has important consequences for the evolution to and within the WR phase, the impact on the overall mass return is small.

The total mass loss is shown in Fig.~\ref{fig16}. This is simply a time-integrated version of Fig.~\ref{fig15}, and its behavior mirrors that of Fig.~\ref{fig15}. The line types are the same as in the previous figure. Despite the differences between the mass-loss rates early and late in the evolution, the total mass lost over the entire stellar life time is rather similar in all sets of evolution models.

Fig.~\ref{fig17} shows the light-to-mass ratio for $L_{\rm{bol}}$, $L_V$, $L_K$. Panels a) and b) are for $Z = 0.020$ and $Z = 0.040$, respectively. Line types are as in Fig.~\ref{fig16}. The model prediction for the bolometric light-to-mass ratio is higher by $\sim 0.2$ dex in the models with {\vrotthree}~{\kms} for both chemical compositions once post-main-sequence stars dominate the light. The fact that there is no difference between the models with $Z = 0.020$ after log(Age[Myr])~=~1.7 is an artifact. This age corresponds to the life time of a 9~{\msun} star, which equals the lowest-mass track available for the new evolution models. After this age, the tracks for all population models are the same. At  $Z = 0.040$, the same effect sets in after about 10~Myr. The increase of the $L/M$ ratio results from the higher luminosities of the new models with rotation. As we showed before, the total mass loss, and therefore the current stellar masses are very similar in the model sets. The increase of the $L/M$ ratio during the post-main-sequence phase in the new models is even more pronounced in the $V$, and in particular in the $K$ band. The passbands favor the contribution by luminous post-main-sequence stars, whereas the bolometric luminosity
tends to have a strong component from hot main-sequence stars whose luminosities are not too different in the three model sets.

$L/M$ for the bolometric luminosity with tracks of {\vrotcero} {\kms} and with the old tracks are essentially identical. Some differences can be recognized for the light-to-mass ratio in the {\em V} and {\em K} bands. There is a higher contribution around $\log Age[Myr] = 0.7$ (Fig.~\ref{fig17}a). In general, the non-rotating tracks evolve faster and are less luminous than models of {\vrotthree}~{\kms}. Therefore, the feature is caused by the earlier excursion to the red part of the HRD in these models.

Fig.~\ref{fig17}b includes the prediction of the models with tracks using a mass-loss rate $\propto Z^{0.5}$ (dot-long dashed line). These tracks are sampled for masses in the range of $40 \le M \le 120$ {\msun}. As a result, their contribution disappears after 10~{\myr}. The models produce significantly higher luminosities when WR stars are present and contribute to the light. This is the case in particular for the $V$ band.

\section{Comparison with observations}

Having compared the new models to the previous generation of non-rotating
models, we are turning to a first comparison with observations. The
incompleteness of the new grid limits this comparison to relatively few
observables. Nevertheless, these early indications can provide guidance for
future modeling efforts.

\subsection{Comparison with synthetic Hess diagrams}

Hess diagrams are useful tools for visualizing actual number densities in HRDs, as opposed to simply plotting evolutionary tracks. This is particularly important for the comparison of stellar life times. Ordinarily, stellar number densities are represented in Hess diagrams. In the following we have generalized this method by plotting integrated population properties such as rotation velocities or abundances weighted by the stellar number densities.

The Hess diagrams were calculated with a grid defined by 56 points over the range $4.9 \ge \log{T_{eff}} \ge 3.5$ and $2.5 \le \log{L/L_\sun} \le 6.5$. Star formation was assumed to be continuous over 100 Myr, which implies star formation and stellar death are in equilibrium for masses as low as $\sim$5~{\msun}. A Salpeter IMF in the range of 1~--~100 M$_\sun$ was used in combination with the new tracks of {\vrotthree} {\kms}. Variables visualized in these diagrams are: star number ($N_{stars}$), current mass ($M_c$), mass-loss rate ($\log{dm/dt}$), mass fraction of H, He, C, N, O, and the rotation velocity. The diagrams were made with an IDL routine that uses the subroutine {\em singlecontour} from the package CHORIZOS \citep{chorizo}.

Stellar statistics from star-forming galaxies in the Local Group such as the Magellanic Clouds, are well suited for comparison with our models. We have used the results of  \citet{venn99}, \citet{venn03}, \citet{crowther02b}, \citet{bouret03}, \citet{hillier03}, \citet{lennon03}, \citet{trundle04}, and \citet{trundle05} who analyzed samples of massive stars in the Small Magellanic Cloud (SMC). The data, including the errors, are shown in Table~\ref{tbl-3}. We compared their nitrogen abundance, current mass, mass-loss rate, and rotation velocity to those parameters in the new Geneva tracks of {\vrotthree} {\kms} via the Hess diagrams.

Fig.~\ref{fig18} shows the Hess-diagram at $Z = 0.004$ for the nitrogen abundance in units of $\log[(N/H)/(N/H)_{\rm ini}]$. For a fair comparison with the data we have re-scaled the nitrogen abundance from the tracks to the average abundance of nitrogen of the SMC as done by \citet{trundle04}. The resulting ZAMS abundance is $12+\log(N/H)_{\rm ini}=6.55$. The grey-scale visualizes the average nitrogen abundance in each temperature-luminosity bin predicted by the models under the assumption of constant star formation. The average value can be the result of the superposition of different evolutionary phases. For instance, an evolutionary track may cross the same bin on its right- and leftward movement, with the star having different chemical composition. In order to facilitate the comparison with the data, we chose matching colors for the symbols representing the data. Each symbol is surrounded by a square. In those cases where the models and data agree very well, the symbol is invisible and only the square appears.

In the case of $Z=0.004$ (as well as $Z=0.008$), the new tracks cover the uppermost mass range only. We substituted the old Geneva tracks for the missing rotating tracks in order to be able to generate the Hess diagram. (This was done in Fig.~\ref{fig18} and all other figures with missing models.) The cut-off luminosity below which the tracks were substituted is indicated by the horizontal line. The data added to Fig.~\ref{fig18} are from Table~\ref{tbl-3}. The average nitrogen abundance for all stars in this table except the A and F supergiants is $12+\log{N/H}\sim7.64$. Very few stars are luminous enough for a comparison with the new models. Taken at face value, those few stars suggest the nitrogen surface abundance is somewhat underestimated by the new stellar models.

In Fig.~\ref{fig19} we show the Hess diagram for the current stellar mass. The ``observed'' masses were determined spectroscopically by \citet{trundle05}. Note that these masses were derived from an atmospheric analysis and not dynamically from binary stars. Therefore the term ``observed'' should be taken with care. The data agree very well with the model predictions, with the caveat that the comparison for most of the stars refers to the old non-rotating tracks. On the basis of the limited sample we conclude there is no discrepancy between evolutionary and spectroscopic masses.

The Hess diagram for the mass-loss rates is in Fig.~\ref{fig20}. The mass-loss rates are of course closely related to the stellar current mass (Fig.~\ref{fig19}) as these rates determine the relation between the masses at any epoch during stellar evolution. The new models match the observation very well in most stars over the range of $-7 < \log{dM/dt} < -6$, with a few stars below this value and some above. Most notably, the predicted rates for B supergiants are higher than observed.

Next we turn to the rotation velocity. Fig.~\ref{fig21} shows the Hess diagram for the rotation velocity $v_{\rm {rot}}$ for the stellar sample in Table~\ref{tbl-3}. The data were corrected by the statistical factor for the distribution of the inclination angle $4/{\pi}$ to obtain $v_{\rm {rot}}$. Therefore the values of the grey-scale are the rotation velocity at the stellar equator. Fig.~\ref{fig21} suggests lower observed rotation velocities than those predicted from the models. The average observed value for massive stars (ignoring A and F supergiants) is $v \sin i \sim 65 \simeq$~{\kms}.

\citet{trundle04} compared their derived rotation velocities with an early set of tracks with rotation from \citet{mame01}. In order to enforce agreement, Trundel et al. had to postulate an average inclination angle of $i = 20\degr$, which means most stars are observed pole-on. Alternatively the observations could be systematically low, or the models too high. Apparently, the discrepancy remains with the new set of tracks, as suggested by Fig.~\ref{fig21}. Trundle et al. discuss a possible overestimate of the rotation velocity in their sample because the line widths could be affected by macroturbulence, which would add to the broadening. However, it seems doubtful that macroturbulence can be sufficiently large to account for the discrepancy. A more likely explanation could be a selection effect: the stars in Trundle et al. and \citet{trundle05} were chosen to have low rotation velocity (Lennon, private communication).

The comparison of the rotation velocities with SMC stars suffers from the incompleteness of the evolutionary models. Since the models with solar chemical composition are complete down to 9~{\msun} (Table~2), we performed a corresponding comparison for the sample of Galactic O stars of \citet{conti77}. We translated the spectral types to temperature and
luminosity by using the \citet{smikal82} calibration. Spectral type bins occupied by more than one star were averaged in $v \sin i$. The 119 stars from the sample produced 33 averaged values in the Hess diagram shown in Fig.~\ref{fig22}. The values are also tabulated in Table~\ref{tbl-4}. Once corrected by $4/{\pi}$, the observed values remain somewhat smaller than the model predictions but the difference is much smaller than that for the SMC stars in  Fig.~\ref{fig21}. Stars with masses lower than 30 {\msun} show a larger discrepancy.

We also compared the rotation velocities of the solar abundance models to those derived by \citet{huang05}. Their sample consists mostly of B stars in 19 young open clusters, which is a useful extension of the previous O-star sample to lower luminosities. As before, we averaged the rotation velocities of identical spectral types. These averaged values are shown in Table~\ref{tbl-5} and plotted in Fig.~\ref{fig23}. Once again, we have corrected the observations by the statistical factor of $4/{\pi}$ to compare with the models. As for the O stars, the models somewhat overestimate the rotation velocities in this collection of Galactic B-type stars.

To summarize, we find good agreement between the models with rotation and the data for the current masses and the mass-loss rates. The agreement for the nitrogen abundances and rotation velocities is poor but the observational material is sparse and subject to selection effects.

\subsection{Red supergiants in the Magellanic Clouds}

One of the main drivers for building a new set of tracks was the long-standing problem of the disagreement between the
predicted and observed RSG number statistics \citep{lanmae95,mame01}. In this section we compare the predictions of the new set of tracks with rotation to samples of RSGs in the Magellanic Clouds.

\citet{maol03} conducted a photometric survey of the Magellanic Clouds using {\em BVR} CCD photometry and found 105 and 158 confirmed RSGs in the SMC and LMC, respectively. These authors performed a comparison of their census with different evolutionary models and found that none of the tested models matched the RSG sequence.
There are two issues involved in this apparent failure: first, the observed and predicted effective temperatures do not match. This mismatch may be mitigated or even completely resolved by adopting a revised, hotter $T_{\rm eff}$ scale. A second, more troublesome issue are the lifetimes. The time spent in the very red part of the HRD is too short in the models. As a result, the stellar number densities in the region where RSGs are observed do not match. However, as pointed out before, most RSGs are produced by stars with masses $< 30$~{\msun} for which we have no evolutionary tracks with rotation at subsolar chemical composition. In this case, our predicted RSG numbers are based on models with the old Geneva tracks only. This leaves open the possibility that the RSG properties are indeed reproduced correctly in both SMC and LMC once the proper evolution models become available. 

Recently, \citet{levesque06} analyzed a new sample of RSGs in the SMC and LMC. Using the same MARCS atmosphere models as in \citet{levesque05} they argue in favor of a 10\% hotter effective temperature scale.  Furthermore, metallicity-dependent systematic differences in the scale are found. Using this new effective temperature scale, the observations match much better the theoretical predictions from the Geneva models as shown in Figs.~\ref{fig24} and \ref{fig25}. 

We have also compared the predictions for models with solar chemical composition to observations of {\em Galactic} RSGs with the new effective temperature scale determined by \citet[][Fig.~\ref{fig26}]{levesque05}. The definition of blue and red supergiants is also indicated in this figure. In this case, models with the new tracks reproduce the regime of Galactic RSGs very well. A simple conclusion from this figure is that most RSGs are produced by stars with initial masses lower than 30 M$_\sun$, explaining the mismatch at lower $Z$ in the SMC and LMC cases. 
Fig.~\ref{fig26} supports the earlier suggestion that a revised effective temperature scale in the Magellanic Clouds could help to solve the mismatch with the new models. However, the state of art of the recent models with rotation is still undergoing revision, and improved models could reach the red part of the observations.

The complexity of the evolutionary models and the incomplete information about different aspects of massive stars do not yet allow us to produce a  more complete set of evolutionary models. \citet{mame01} produced a set of tracks of {\vrotthree} {\kms} with $Z = 0.004$. These tracks were specifically built with the purpose to reach the RSG zone. In contrast, the tracks used in the present work were produced with the goal of reproducing the overall properties of all massive stars in the upper HRD. In particular, modeling the behavior of WR stars was a major objective. In order to simplify the calculations, convection in the envelope was performed using a density scale height instead of a pressure scale height. This choice affects the temperatures in the RSG phase, making the stars hotter. The very good agreement with Galactic RSG favors this choice at solar chemical composition.

\subsection{Observable star number ratios}

Reproducing the duration of individual stellar phases, such as the O, WR, or RSG phase, is a major quest in stellar evolution modeling \citep{mema03,mema05}. In this section we use the predicted ratios of Section~3. We assume that the star numbers are from a stable population, i.e., star formation has been ongoing for a sufficiently long interval to allow the relevant stellar phases to reach equilibrium. The input parameters are the same as those described in Section~3.

\subsubsection{WC/WN}

Fig.~\ref{fig27} shows the WC/WN ratio for different stellar systems from \citet{maolpa03} for the SMC, \citet{breysa99} for the LMC, \citet{hadfield07} for the solar neighborhood, \citet{crockett06} for M~31, \citet{hadfield05} for M~83, and for IC~10 by \citet{mashol02}. We have included in the figure three models with different set of tracks: old Geneva (solid line), new tracks of {\vrotthree} {\kms} (dotted and dot-dashed lines), and new tracks of {\vrotcero} {\kms} (short-dashed line). For these models we have adopted a temperature limit to enter to the WR phase of $\log T_{\eff} = 4.0$, and we have used a temperature limit of $\log T_{\eff} = 4.4$ just for the new tracks of {\vrotthree} {\kms} (dot-short dashed line).

As shown in Section~3, very high values for WC/WN are obtained for models using the old Geneva tracks and a higher temperature limit of $\log T_{\eff} = 4.4$ for the WR phase. The previous values for this ratio are higher than the upper limit of the plot in Fig.~\ref{fig27}. In contrast, the new tracks with rotation do no longer overestimate the WC/WN ratio at low $Z$. At high $Z$, however, they seem to underestimate this ratio. Note that the {\em dotted line} fits better the trend shown by the data, especially for low values of the oxygen abundance.

Models with the new tracks of {\vrotthree} {\kms} predict a higher number of WN stars for solar and higher oxygen abundance than expected from the observations. As discussed in the previous sections, the number of WNs has increased because (i) the most massive stars enter the WR phase earlier and (ii) this phase lasts longer due to a combination of rotation and mass loss (in particular at higher metallicities where these effects are more important). Models with tracks of {\vrotcero} {\kms} predict a ratio in better agreement with the observation at solar composition. An alternative way to resolve this apparent paradox has been suggested by \citet{mema05}, who argued in favor of an underestimate of the mass loss during the WNL phase.

The WC/WN ratio for IC~10 is still unclear since the statistics was based on unconfirmed WR candidates, but seems to be in the range of the model prediction. The Milky Way data were based upon stars within galacto-centric distances $<7$ kpc, 7-10 kpc, and $>10$ kpc.

\subsubsection{WR/O}

An important observational constraint for stellar evolution models is provided by the variation of the number ratio of WR to O-type stars with metallicity. Fig.~\ref{fig28} shows the WR/O ratio for different stellar populations taken from \citet{mame94} at different radius from the Sun. The value for the SMC agrees with recent results with $\sim 1000$ O-type stars \citep{evans04} versus 12 WR \citep{maolpa03} giving a WR/O $= 0.01$. The data are compared to models with the old Geneva tracks (solid line), new tracks of {\vrotthree} {\kms} (dotted line), and new tracks of {\vrotcero} {\kms} (short-dashed line).

Symbols annotated with an N for the SMC and LMC are based on recent determinations of the oxygen abundances for these two galaxies. For the SMC, \citet{lennon03} found a value of $12+\log(O/H) = 8.15$ in B stars. \citet{cos00} derived an oxygen abundance of 8.22 in a sample of planetary nebulae. In this work we have adopted the oxygen abundance obtained when using a value of 8.75 for the solar composition and scaling this value to $Z=0.004$ (i.e., dividing by a factor of 5). We did the same for the LMC oxygen abundance. The fit to the observations by our models becomes slightly better if we adopted a lower solar oxygen abundance, i.e., shifting the zero point of the models along the horizontal axis. 

Recent determinations of the solar oxygen abundances give values significantly lower \citep[$12+\log(O/H) = 8.69;$][]{all02,asplun05} than the one of \citet{anders89}.
The dot-long dashed line shows the same values for the WR/O ratio but considering lower values for the oxygen abundance for solar and higher chemical composition in Fig.~\ref{fig28}. 

Obviously, the real situation is more complex than implied by our simple scaling, in particular since the Fe abundance is the more important parameter driving evolution. Nevertheless, there seems to be a lack of WR stars in the models for solar and higher chemical composition.

\subsubsection{R/WR}

Fig.~\ref{fig29} shows our models for the ratio of RSGs over WR stars (R/WR) using the same layout as in Fig.~\ref{fig27}. Symbols are the observed ratios from \citet{mas02}. The figure includes only RSGs that lie in the same areas for which surveys are complete for WRs and follow the definition of a RSG as described in Section~3.2.3.

Models using the old Geneva tracks (solid line) do not match the observations, except for the LMC oxygen abundance. The new tracks of {\vrotthree} {\kms} do not include tracks for masses $< 30$ {\msun} for oxygen abundance less than solar. The mass range above 30~{\msun} is too small to produce a significant number of RSGs with the new tracks. Therefore we generate the same theoretical values as those obtained with the old Geneva tracks. For this reason, we have truncated the graphs below solar oxygen abundance. 

For solar and higher oxygen abundances, the new models and the data disagree. This is the result of too few WR stars being predicted by the models. The adopted temperature limit for the WR phase does not significantly affect the results. This can be seen  as seen for the models with tracks of {\vrotthree} {\kms} (dot-short dashed line). Recent results for M~33 and M~31 suggest lower oxygen abundances in H{\small II} regions by about 0.1 dex \citep{maolpa03,crockett06}. The lower abundances do not significantly affect our conclusions but should be kept in mind when inspecting Fig.~\ref{fig29}.

\subsubsection{B/R}

The final star number ratio to compare to our models is the blue-to-red supergiant ratio. 
Fig.~\ref{fig30} shows the B/R ratio for different sets of data as a function of the oxygen abundance. There are several determinations for this ratio in the Magellanic Clouds. We used the homogeneous survey of stars in the SMC and LMC by \citet[triangles]{mas02} and \citet[squares]{maol03} for our comparison.

\citet{maol03} discuss contamination effects for K and M-type stars in surveys when a color cut-off of ($V - R$) $> 0.6$ is imposed. Foreground contamination can significantly affect the observed numbers. $11\%$ and $5.3\%$ of the bona fide RSGs seen towards the SMC and LMC, respectively, are foreground objects in the survey of \citet{maol03}. Another error source is the calibration of the relation to transform colors into effective temperature. 

We also included in Fig.~\ref{fig30} the results of the study of \citet{lanmae95} who analyzed different samples of blue and red supergiants in associations and in extragalactic fields. The trend of these observations suggests a lower value of B/R for lower chemical composition and higher values for higher chemical composition. This result disagrees with the numbers of \citet{mas02} and \citet{maol03}.

The third data set in Fig.~\ref{fig30} has been obtained by \citet{huevo02}. Their fit was derived by using a sample of 45 clusters in the Milky Way and 4 clusters in Magellanic Clouds.  To obtain the long-dashed line in the plot, we assumed a value of B/R~=~30 at solar composition, which is similar to the value obtained by \citet{huevo02} for Galactic clusters.

As before, three models are shown in Fig.~\ref{fig30}, with the line types as in Fig.~\ref{fig28}. We do not plot the B/R ratio for chemical composition lower than solar for the same reason as in the previous subsection. Models with the old Geneva tracks do not fit the observations, whereas models with the new tracks of {\vrotthree} {\kms} match the observations for oxygen abundance higher than solar. Models with the new tracks of {\vrotcero} {\kms} predict similar values as the models with the old Geneva tracks. As argued before, this ratio does not depend on the minimum temperature for the WR phase.

It is evident that the data are very inconsistent for subsolar chemical composition. Nevertheless, the data agree in a trend of higher B/R at higher O/H. Our models predict a similar trend. However, our predictions are still affected by
uncertainties in the zero point for the oxygen abundance. As in the case of the WR/O ratio, our values agree better with the observations if we adopt a lower solar oxygen abundance as shown by the dot-long dashed line in Fig.~\ref{fig30}.

\section{Conclusions}

Progress in stellar evolution modeling is a rather discontinuous process when viewed from a ``user's'' perspective. Population synthesis modelers benefit from and are at the mercy of updates and new releases by the major stellar evolution modeling groups. Evolution models for massive stars have undergone dramatic improvements over the past several decades. Major milestones were set with the realization of the importance of stellar mass loss, of convective processes, and of the influence of chemical composition. Each time the release of a new set of stellar evolution models led to sometimes major revisions of the predictions by population synthesis models.

The latest major release of revised evolution models by the Geneva and Padova groups occurred in the 1990s when new opacities were included and the importance metallicity-dependent mass loss was highlighted. These sets of evolution models are currently employed by virtually all population synthesis code in use. Now we are in an epoch of another quantum leap in massive star evolution. The importance of stellar rotation for massive stars has been recognized and evolution calculations have been done over the past years. Only very recently has a grid of evolution models with rotation been completed that is sufficiently broad and dense to be suitable for implementation. This model grid, with ZAMS velocities of 0 and 300~{\kms}, has been implemented into Starburst99.

We acknowledge the shortcomings of the models presented in this paper. The limited mass coverage of the available stellar models often introduces serious biases making the comparison with observations difficult. Nevertheless, some of the trends we have found in our study are significant improvements to the stellar populations study.

Massive hot, blue stars tend to be more luminous and hotter when rotation is accounted for. This results from the increased convective core and larger amount of mixing reducing the surface opacities. When coupled with synthesis models, these properties lead to significantly different population parameters. For a fixed stellar mass, a young population with rotating stars becomes more luminous at almost all wavelengths of interest. Specifically, the ultraviolet to near-infrared luminosities can increase by factors of order 2, in particular when RSGs are present. If such models are compared with observed light-to-mass ratios of young star clusters, the derived mass may become smaller by the same factor in comparison with the results of earlier studies. This has important consequences for our understanding of the low-mass end of the stellar initial mass function.

The impact of the new models at wavelengths below 912~{\AA} is even more dramatic. The SED in the extreme ultraviolet becomes enhanced and harder when compared with previous synthesis models. The effect is most pronounced in the ionized helium continuum below 228~{\AA} due to the increased WR star contribution. Even at shorter wavelengths in the neutral helium and hydrogen continua below 504 and 912~{\AA}, respectively, the changes are significant. Since the Lyman continuum flux increases by a factor of 2 to 3, the derived stellar masses or star-formation rates from hydrogen recombination will decrease by the same factor. A large fraction of the increased photon output is provided by WR stars whose frequency is strongly dependent on the chemical composition. Therefore the prediction recombination-line fluxes and star-formation tracers become more dependent on $Z$ than in previous models. 

Given the potential impact of these new predictions it becomes crucial to perform a critical assessment of the 
additional assumptions that were made. The increase and hardening of the spectrum is the result of the stellar interior and atmosphere connection, which rests on the definition of the effective temperature and radius of the star. Moreover, atmosphere models for WR stars must be adopted. Another major area of concern are the mass-loss rates, in particular during the WR phase. The mass loss rates used in the default version of the new tracks with rotation do not include a metallicity dependence of the rates. It is also a matter of debate how the competing effects and mass loss affect the stellar properties. 

Nevertheless, this study should be viewed as a first step in the development of more realistic and accurate models for stellar populations. Should our initial results be confirmed when more mature evolutionary tracks are available, several widely used star-formation and initial mass function indicators will have to be revised.

\acknowledgments

Thanks to Leonardo Ubeda who helped with the Hess diagrams. Support for this work was provided by NASA through grant O1166 from the Space Telescope Science Institute, which is operated by the Association of Universities for Research in Astronomy, Inc., under NASA contract NAS5-26555.

\appendix

\clearpage

\begin{deluxetable}{lrrrr}
\tabletypesize{\scriptsize}
\tablecaption{Minimum mass limits for entering the WR phase for different sets of tracks (in {\msun}). 
\label{tbl-1}}
\tablewidth{0pt}

\tablehead{
\colhead{} & \multicolumn{4}{c}{Chemical Composition } \\
\colhead{Mass limit} & 
\multicolumn{1}{c}{$Z=0.004$} & 
\multicolumn{1}{c}{$Z=0.008$} & 
\multicolumn{1}{c}{$Z=0.02$} & 
\multicolumn{1}{c}{$Z=0.04$}\\
}

\startdata

$M_{\rm{WR}}$\tablenotemark{a}({\vrotthree} {\kms}) &    32 &      25 &      22 &      21 \\
$M_{\rm{WR}}$({\vrotcero} {\kms}) & \nodata\tablenotemark{b} & \nodata &      37 &      29 \\
$M_{\rm{OWR}}\tablenotemark{a}$({\vrotthree} {\kms}) &    75 &      69 &      45 &      39 \\
$M_{\rm{OWR}}$({\vrotcero} {\kms}) & \nodata & \nodata &      62 &      42 \\
$M_{\rm{WR}}$(HML\tablenotemark{c}) &    42 &      35 &      25 &      21 \\
$M_{\rm{WR}}$(LML\tablenotemark{c}) &    52 &      42 &      32 &      25 \\


\enddata

\tablenotetext{a}{$M_{\rm{OWR}}$. The mass limit to enter to the WR phase on the main-sequence. $M_{\rm{WR}}$. The mass limit to enter the WR phase at any time during the evolution of the star \citep{mema03,mema05}.}
\tablenotetext{b}{No computations are available with {\vrotcero} {\kms}}
\tablenotetext{c}{Old Geneva tracks denoted as HML and LML for high and normal mass-loss rates, respectively.}
\end{deluxetable}


\begin{deluxetable}{rrrr}
\tabletypesize{\scriptsize}
\tablecaption{Mass range (in {\msun}) for the new models with rotation. \label{tbl-2}}
\tablewidth{0pt}

\tablehead{
\multicolumn{4}{c}{Tracks with initial {\vrotthree} {\kms}}\\ 
\colhead{$Z=0.004$} & 
\multicolumn{1}{c}{$Z=0.008$} & 
\multicolumn{1}{c}{$Z=0.02$} & 
\multicolumn{1}{c}{$Z=0.04$}\\
}
\startdata
    120 &     120 &     120 &     120 \\
\nodata & \nodata &      85 &      85 \\
     60 &      60 &      60 &      60 \\
     40 &      40 &      40 &      40 \\
     30 &      30 & \nodata & \nodata \\
\nodata & \nodata &      25 &      25 \\
\nodata & \nodata &      20 &      20 \\
\nodata & \nodata &      15 & \nodata \\
\nodata & \nodata &      12 & \nodata \\
\nodata & \nodata &       9 & \nodata \\

\tableline

\multicolumn{4}{c}{Tracks with initial {\vrotcero} {\kms}} \\

\colhead{$Z=0.004$} & 
\multicolumn{1}{c}{$Z=0.008$} & 
\multicolumn{1}{c}{$Z=0.02$} & 
\multicolumn{1}{c}{$Z=0.04$}  \\

\tableline

\nodata & \nodata &     120 &     120 \\
\nodata & \nodata &      85 & \nodata \\
\nodata & \nodata &      60 &      60 \\
\nodata & \nodata &      40 & \nodata \\
\nodata & \nodata & \nodata & \nodata \\
\nodata & \nodata &      25 &      25 \\
\nodata & \nodata &      20 & \nodata \\
\nodata & \nodata &      15 & \nodata \\
\nodata & \nodata &      12 & \nodata \\
\nodata & \nodata &       9 & \nodata \\

\tableline

\multicolumn{4}{c}{Tracks with metallicity dependent WR rates}\\

\colhead{$Z=0.004$} & 
\multicolumn{1}{c}{$Z=0.008$} & 
\multicolumn{1}{c}{$Z=0.02$} & 
\multicolumn{1}{c}{$Z=0.04$}  \\

\tableline

\nodata & \nodata & \nodata &     120 \\
\nodata & \nodata & \nodata &      85 \\
\nodata & \nodata & \nodata &      60 \\
\nodata & \nodata & \nodata &      40 \\
\nodata & \nodata & \nodata & \nodata \\
\nodata & \nodata & \nodata & \nodata \\
\nodata & \nodata & \nodata & \nodata \\
\nodata & \nodata & \nodata & \nodata \\
\nodata & \nodata & \nodata & \nodata \\
\nodata & \nodata & \nodata & \nodata \\


\enddata

\end{deluxetable}


\begin{deluxetable}{lcccccc}
\tabletypesize{\scriptsize}
\tablecaption{SMC observations for comparison with $Z=0.004$ models. 
\label{tbl-3}}
\tablewidth{0pt}

\tablehead{
\colhead{Name} &
\colhead{$\log{T}_{\eff}$} &
\colhead{$\log{L/L_\sun}$} &
\colhead{$12+\log{N/H}$} &
\colhead{$V\sin{i}$} &
\colhead{Mass} &
\colhead{dM/dt} 
 \\
\colhead{} &
\colhead{} &
\colhead{} &
\colhead{} &
\colhead{[km s$^{-1}$]} &
\colhead{[M$_\sun$]} &
\colhead{[$10^{-6}$M$_\sun$yr$^{-1}$]} 
 \\
 }
\startdata
\multicolumn{7}{c}{With UVES at VLT \citet{trundle04}} \\
AV215   &   4.431   &  5.63 $\pm$ 0.06   &    7.94 $\pm$ 0.10   &     90  &  26    &    1.3500 $\pm$ 0.14\\
AV104   &   4.439   &  5.31 $\pm$ 0.06   &    7.40 $\pm$ 0.13   &     79  &  19    &    0.4000 $\pm$ 0.05\\
AV216   &   4.415   &  5.00 $\pm$ 0.06   &    7.63 $\pm$ 0.19   &     75  &  36    &    0.1200 $\pm$ 0.02\\
AV362   &   4.146   &  5.50 $\pm$ 0.06   &    8.20 $\pm$ 0.25   &     50  &  17    &    0.8000 $\pm$ 0.08\\
AV22    &   4.161   &  5.04 $\pm$ 0.06   &    7.90 $\pm$ 0.14   &     46  &   8    &    0.2300 $\pm$ 0.02\\
AV18    &   4.279   &  5.44 $\pm$ 0.06   &    7.50 $\pm$ 0.31   &     49  &  17    &    0.2300 $\pm$ 0.03\\
AV210   &   4.312   &  5.41 $\pm$ 0.06   &    7.60 $\pm$ 0.12   &     65  &  15    &    0.2000 $\pm$ 0.05\\
Sk191   &   4.352   &  5.77 $\pm$ 0.06   &    7.63 $\pm$ 0.20   &     95  &  33    &    0.6800 $\pm$ 0.07\\
\multicolumn{7}{c}{EMMI at the NTT \citet{trundle05}} \\
AV78    &   4.332   &  5.92 $\pm$ 0.06   &    8.30 $\pm$ 0.26   &   \nodata &  57    &    2.2900 $\pm$ 0.34\\
AV264   &   4.352   &  5.44 $\pm$ 0.06   &    7.88 $\pm$ 0.12   &   \nodata &  16    &    0.2900 $\pm$ 0.06\\
AV242   &   4.398   &  5.67 $\pm$ 0.06   &    7.39 $\pm$ 0.19   &   \nodata &  35    &    0.8400 $\pm$ 0.13\\
AV420   &   4.431   &  5.35 $\pm$ 0.06   &    7.44 $\pm$ 0.12   &   \nodata &  19    &    0.3400 $\pm$ 0.15\\
AV151   &   4.204   &  5.28 $\pm$ 0.06   &    7.55 $\pm$ 0.18   &   \nodata &  15    &    0.1600 $\pm$ 0.04\\
AV56    &   4.217   &  5.88 $\pm$ 0.06   &    7.93 $\pm$ 0.22   &   \nodata &  38    &    0.5100 $\pm$ 0.08\\
AV443   &   4.217   &  5.79 $\pm$ 0.06   &    8.26 $\pm$ 0.33   &   \nodata &  30    &    0.4500 $\pm$ 0.09\\
AV10    &   4.230   &  5.21 $\pm$ 0.06   &    7.69 $\pm$ 0.27   &   \nodata &  13    &    0.1500 $\pm$ 0.03\\
AV373   &   4.279   &  5.42 $\pm$ 0.06   &    7.46 $\pm$ 0.00   &   \nodata &  16    &    0.1600 $\pm$ 0.04\\
AV96    &   4.342   &  5.39 $\pm$ 0.06   &    7.69 $\pm$ 0.24   &   \nodata &  15    &    0.2400 $\pm$ 0.06\\
\multicolumn{7}{c}{O-type dwarfs from \citet{bouret03}} \\
MPG 355 &   4.72   &   6.03              &   7.95               &  110   &   65    &    2.5100      \\
MPG 324 &   4.62   &   5.52              &   7.27               &   70   &   40    &    0.2500      \\
MPG 368 &   4.60   &   5.42              &   7.75               &   60   &   40    &    0.1600      \\
MPG 113 &   4.60   &   5.13              &   7.27               &   36   &   30    &    0.0030      \\
MPG 487 &   4.54   &   5.13              &   6.41               &   20   &   25    &    0.0030      \\
MPG 12  &   4.48   &   4.93              &   7.95               &   60?  &   20    &    0.0001      \\
\multicolumn{7}{c}{O-type supergiants from \citet{hillier03}} \\
\nodata &   4.53   &   5.60              &   6.30               &   60    &   \nodata   &   \nodata   \\
\nodata &   4.52   &   5.53              &   8.39               &   70    &   \nodata   &   \nodata   \\
\multicolumn{7}{c}{AF-type supergiants from \citet{venn99} and \citet{venn03}} \\
\nodata &   3.98   &   4.77              &   7.79               &   25    &   \nodata   &   \nodata   \\
\nodata &   3.975  &   4.59              &   7.70               &   15    &   \nodata   &   \nodata   \\
\nodata &   3.93   &   4.47              & \nodata              &   25    &   \nodata   &   \nodata   \\
\nodata &   3.92   &   5.18              &   7.72               &   20    &   \nodata   &   \nodata   \\
\nodata &   3.91   &   4.85              &   7.88               &   15    &   \nodata   &   \nodata   \\
\nodata &   3.905  &   4.62              &   7.07               &   25    &   \nodata   &   \nodata   \\
\nodata &   3.905  &   4.87              &   7.29               &   20    &   \nodata   &   \nodata   \\
\nodata &   3.895  &   4.98              &   7.56               &   20    &   \nodata   &   \nodata   \\
\nodata &   3.875  &   4.58              &   7.80               &   20    &   \nodata   &   \nodata   \\
\nodata &   3.86   &   4.48              &   6.80               &   36    &   \nodata   &   \nodata   \\


\enddata


\end{deluxetable}

\clearpage
\begin{deluxetable}{ccr}
\tabletypesize{\scriptsize}
\tablecaption{Average deprojected rotational velocities derived from the data of \citet{conti77}. 
\label{tbl-4}}
\tablewidth{0pt}

\tablehead{
\colhead{$T_{\eff}$} & \colhead{$L/L_\sun$} & \colhead{\^{\em v}$_{\rm rot}$} \\
}
\startdata

     4.445  &  4.460  &      97\\
     4.485  &  5.630  &    104\\
     4.525  &  5.435  &    107\\
     4.505  &  5.340  &    108\\
     4.420  &  4.815  &    108\\
     4.415  &  5.415  &    112\\
     4.535  &  5.790  &    112\\
     4.515  &  5.725  &    116\\
     4.465  &  5.570  &    119\\
     4.570  &  5.905  &    131\\
     4.610  &  5.905  &    134\\
     4.555  &  5.585  &    144\\
     4.580  &  5.730  &    153\\
     4.645  &  6.205  &    153\\
     4.525  &  5.760  &    159\\
     4.630  &  5.760  &    164\\
     4.680  &  5.995  &    166\\
     4.465  &  5.040  &    166\\
     4.540  &  5.530  &    167\\
     4.500  &  4.850  &    175\\
     4.535  &  5.110  &    175\\
     4.485  &  5.190  &    179\\
     4.650  &  5.900  &    180\\
     4.485  &  4.715  &    181\\
     4.545  &  5.820  &    193\\
     4.615  &  5.625  &    208\\
     4.580  &  5.415  &    221\\
     4.570  &  5.645  &    229\\
     4.600  &  5.815  &    238\\
     4.555  &  5.230  &    243\\
     4.520  &  4.985  &    251\\
     4.600  &  5.520  &    268\\
     4.570  &  5.325  &    270\\

\enddata

\end{deluxetable}

\clearpage
\begin{deluxetable}{ccrccrccr}
\tabletypesize{\scriptsize}
\tablecaption{Same as in Table~\ref{tbl-4} but derived from the data of \citet{huang05}. 
\label{tbl-5}}
\tablewidth{0pt}

\tablehead{
\colhead{$T_{\eff}$} & \colhead{$L/L_\sun$} & \colhead{\^{\em v}$_{\rm rot}$} &
\colhead{$T_{\eff}$} & \colhead{$L/L_\sun$} & \colhead{\^{\em v}$_{\rm rot}$} &
\colhead{$T_{\eff}$} & \colhead{$L/L_\sun$} & \colhead{\^{\em v}$_{\rm rot}$} \\
}
\startdata

  4.400  & 4.500 &      1 & 4.250  & 2.750 &   131 & 4.175  & 2.500 &   192 \\
  4.350  & 4.250 &      6 & 4.150  & 3.000 &   132 & 4.200  & 2.750 &   193 \\
  4.375  & 4.750 &      8 & 4.250  & 3.000 &   136 & 4.400  & 3.875 &   199 \\
  4.275  & 4.000 &    14 & 4.350  & 3.625 &   136 & 4.325  & 3.625 &   202 \\
  4.325  & 4.375 &    23 & 4.450  & 4.375 &   139 & 4.250  & 3.250 &   203 \\
  4.200  & 3.250 &    29 & 4.200  & 2.625 &   140 & 4.450  & 4.500 &   205 \\
  4.300  & 3.000 &    31 & 4.475  & 4.125 &   144 & 4.175  & 2.750 &   208 \\
  4.225  & 2.750 &    36 & 4.100  & 2.875 &   144 & 4.375  & 3.875 &   213 \\
  4.150  & 3.250 &    39 & 4.325  & 3.500 &   145 & 4.250  & 3.125 &   213 \\
  4.450  & 4.625 &    41 & 4.450  & 4.125 &   146 & 4.375  & 4.125 &   214 \\
  4.225  & 3.375 &    45 & 4.350  & 3.125 &   146 & 4.250  & 3.375 &   218 \\
  4.050  & 2.750 &    47 & 4.325  & 3.250 &   149 & 4.225  & 3.125 &   222 \\
  4.375  & 3.500 &    47 & 4.400  & 4.250 &   151 & 4.275  & 3.250 &   223 \\
  4.425  & 4.375 &    52 & 4.150  & 2.500 &   153 & 4.225  & 3.500 &   228 \\
  4.300  & 4.000 &    60 & 4.425  & 3.875 &   155 & 4.225  & 2.875 &   230 \\
  4.150  & 2.750 &    62 & 4.350  & 4.375 &   156 & 4.475  & 4.750 &   232 \\
  4.125  & 2.500 &    66 & 4.125  & 2.875 &   159 & 4.275  & 3.375 &   236 \\
  4.250  & 2.625 &    74 & 4.200  & 3.000 &   159 & 4.225  & 3.000 &   240 \\
  4.425  & 3.750 &    78 & 4.400  & 4.125 &   159 & 4.325  & 3.750 &   253 \\
  4.300  & 3.250 &    86 & 4.275  & 3.000 &   162 & 4.325  & 4.000 &   255 \\
  4.250  & 3.500 &    88 & 4.350  & 3.875 &   162 & 4.425  & 4.625 &   255 \\
  4.375  & 3.625 &    91 & 4.275  & 3.125 &   163 & 4.275  & 3.500 &   263 \\
  4.200  & 3.125 &    93 & 4.425  & 4.250 &   167 & 4.175  & 2.625 &   269 \\
  4.425  & 4.500 &    93 & 4.350  & 3.750 &   167 & 4.225  & 2.625 &   271 \\
  4.425  & 4.000 &    94 & 4.400  & 3.750 &   170 & 4.200  & 2.500 &   271 \\
  4.175  & 3.000 &    95 & 4.050  & 2.625 &   171 & 4.325  & 3.875 &   277 \\
  4.300  & 3.375 &   102 & 4.300  & 3.625 &  171 & 4.350  & 4.000 &   278 \\
  4.350  & 3.375 &   103 & 4.425  & 4.125 &  171 & 4.300  & 3.750 &   281 \\
  4.325  & 4.250 &   106 & 4.375  & 3.750 &  172 & 4.275  & 3.625 &   292 \\
  4.475  & 4.875 &   110 & 4.350  & 3.500 &  172 & 4.375  & 4.000 &   293 \\
  4.150  & 3.500 &   118 & 4.400  & 4.375 &  172 & 4.325  & 4.500 &   297 \\
  4.450  & 4.000 &   122 & 4.300  & 3.500 &  172 & 4.200  & 2.875 &   309 \\
  4.400  & 4.000 &   126 & 4.250  & 2.875 &  173 & 4.275  & 3.750 &   326 \\
  4.350  & 4.125 &   127 & 4.475  & 4.625 &  180 & 4.050  & 2.500 &   351 \\
  4.450  & 4.250 &   129 & 4.375  & 4.375 &  183 & 4.075  & 3.000 &   397 \\
  4.150  & 2.625 &   130 & 4.325  & 3.375 &  189 & 4.075  & 2.500 &   402 \\


\enddata

\end{deluxetable}

\clearpage

\begin{figure}
\plotone{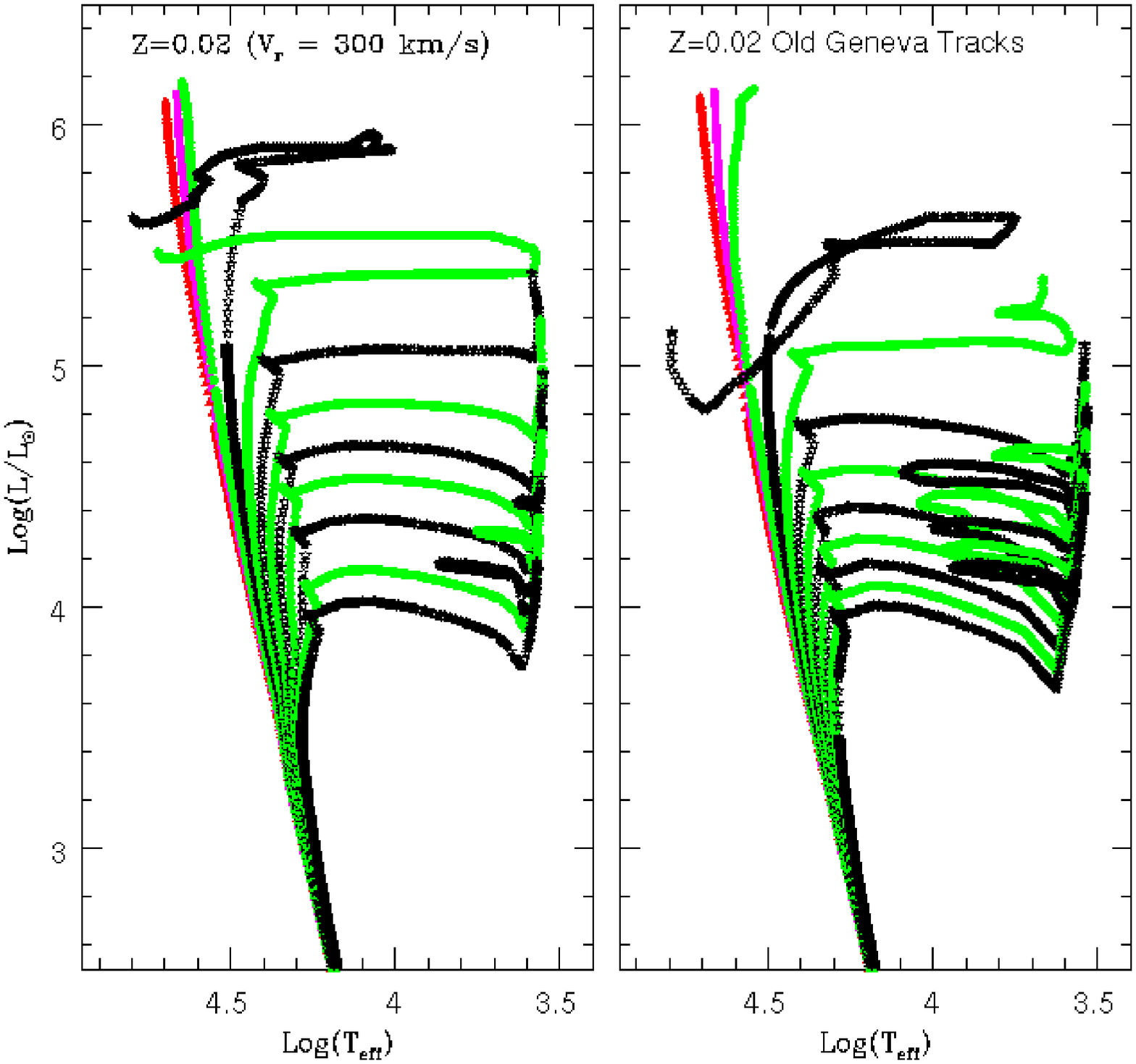}
\end{figure}


\begin{figure}
\plotone{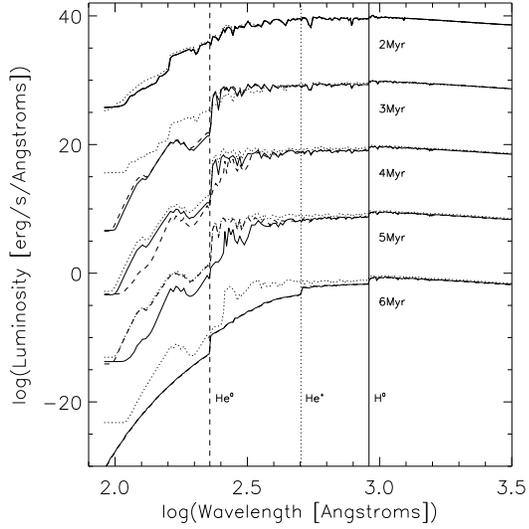}
\caption{SEDs for young populations with solar composition and $10^6$~{\msun}. Solid: previous tracks with enhanced mass loss; short-dashed: {\vrotcero}; dotted: {\vrotthree} {\kms}. For clarity, the SEDs for 3, 4, 5, and 6 Myr are displaced by -10, -20, -30 and -40 dex from the original value, respectively. 
\label{fig3}}
\end{figure}


\begin{figure}
\plotone{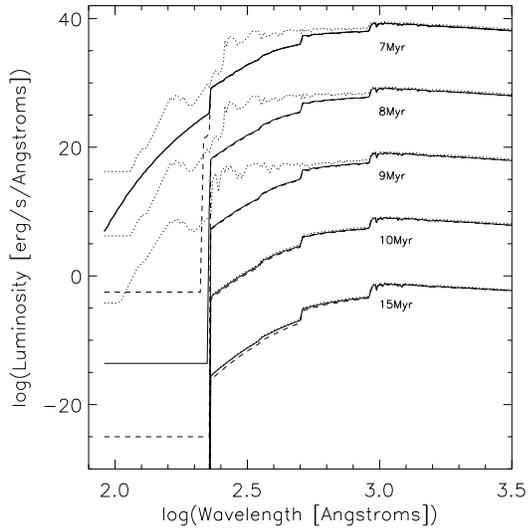}
\caption{Same as Fig.~\ref{fig3}, but for ages of 7, 8, 9, 10, and 15~{\myr}, displaced by -10, -20, -30, and -40 dex, respectively, except for the model at 7~{\myr} which is unshifted.
\label{fig4}}
\end{figure}


\begin{figure}
\plotone{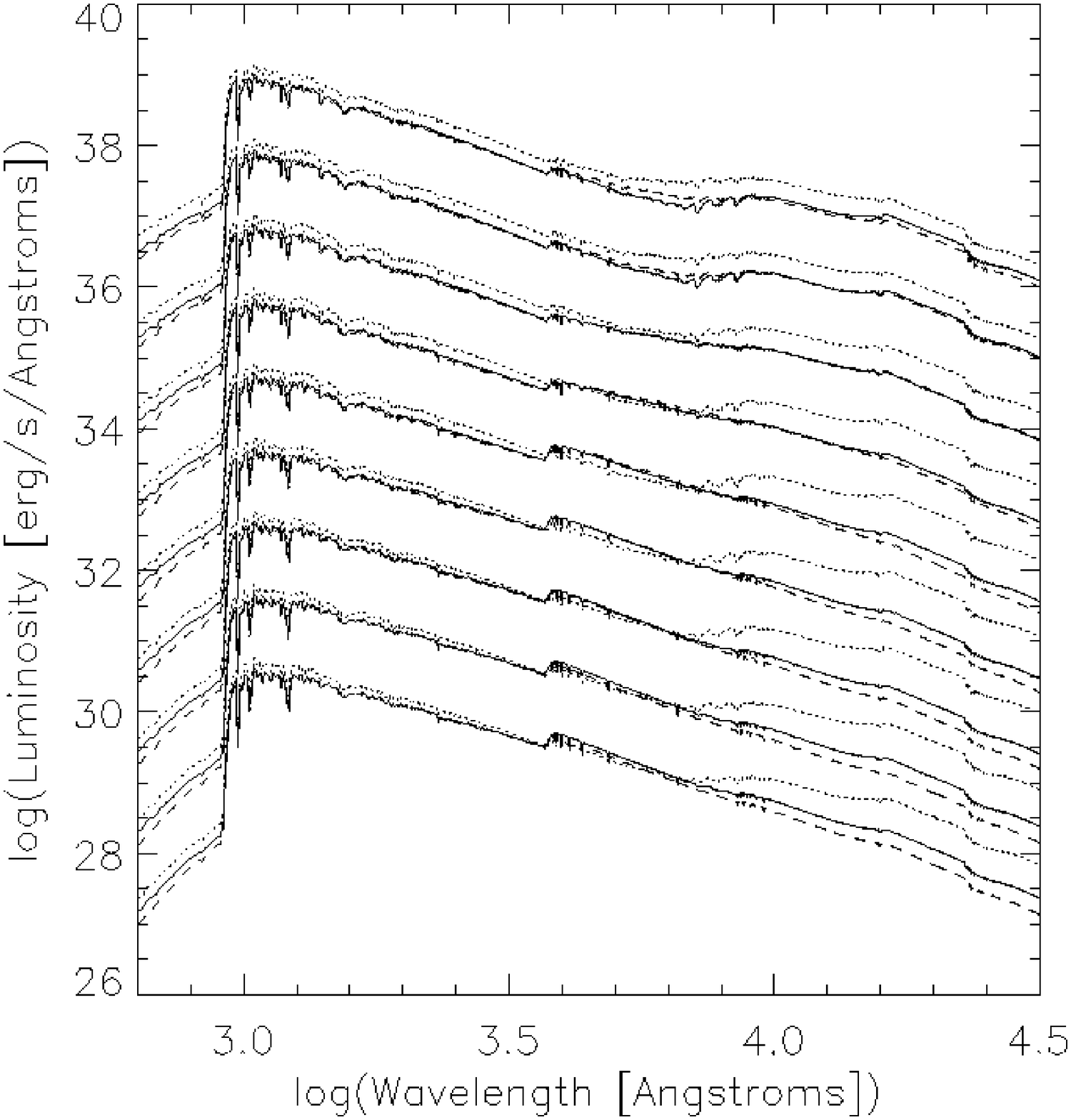}
\caption{Red part of the SEDs for ages 11~--~19 {\myr} (top to bottom) and displaced incrementally by -1.0~dex from the original luminosity, except for the 11~Myr old model. Line types as in Fig.~\ref{fig3}.
\label{fig5}}
\end{figure}


\begin{figure}
\plotone{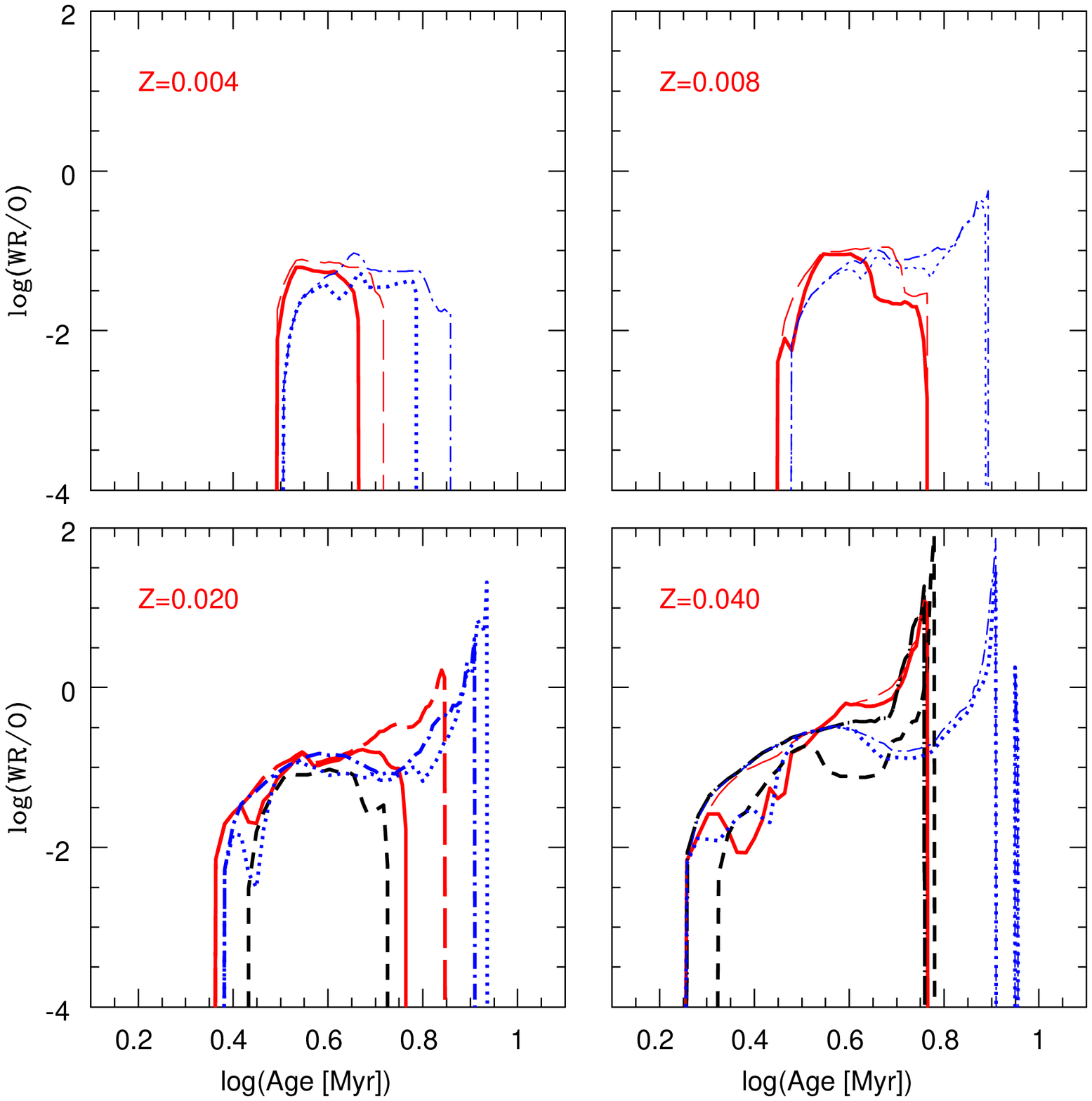}
\caption{Ratio of WR over O stars for an instantaneous burst. Same line types as in the previous figures: $\log{T_{\eff}}\ge4.4$ (thick solid and dotted lines) and $\log{T_{\eff}}\ge4.0$ (long-dashed, dot-dashed and thick short-dashed lines) for models with the old Geneva tracks, new tracks with {\vrotthree} and {\vrotcero} {\kms}, respectively. Thick black dot-long-dashed line: models with new tracks for different WR mass-loss rates and $\log{T_{\eff}} \ge 4.0$ at $Z=0.040$.
\label{fig6}}
\end{figure}


\begin{figure}
\plotone{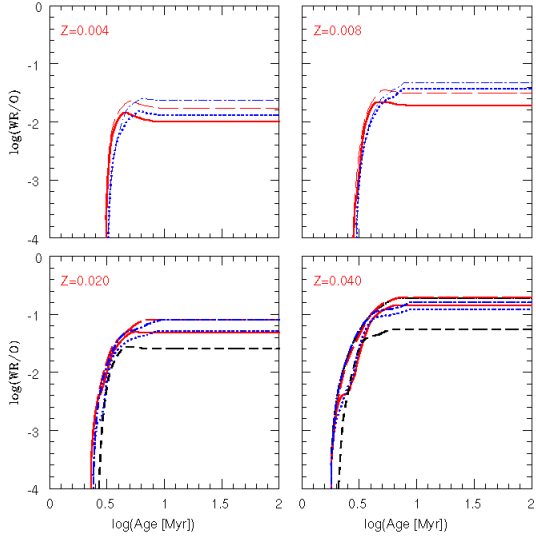}
\caption{Same as Fig.~\ref{fig6}, but for continuous star formation.
\label{fig7}}
\end{figure}


\begin{figure}
\plotone{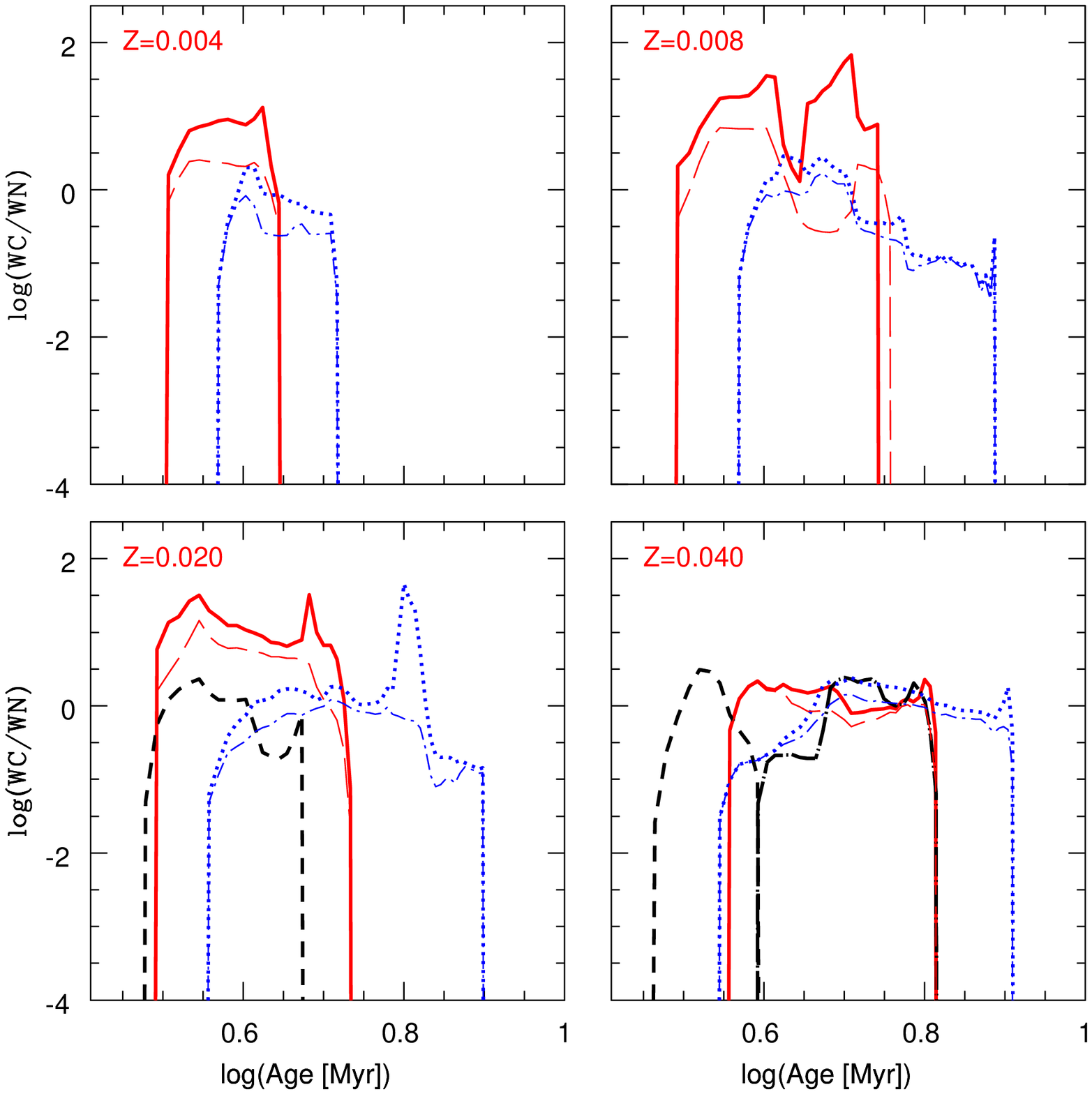}
\caption{Same as Fig.~\ref{fig6}, but for the ratio of WC over WN stars.
\label{fig8}}
\end{figure}


\begin{figure}
\plotone{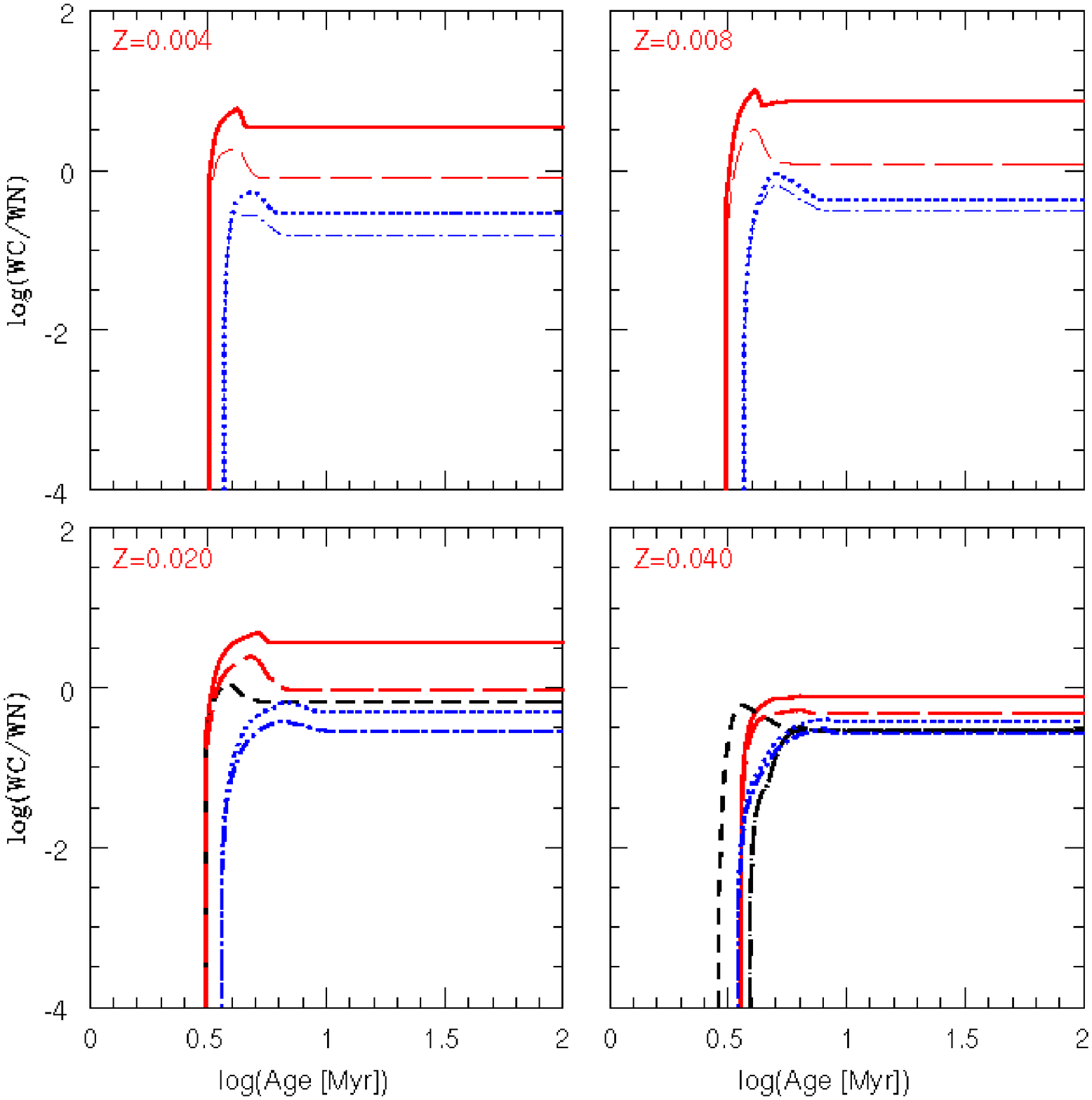}
\caption{Same as Fig.~\ref{fig7}, but for the ratio of WC over WN stars.
\label{fig9}}
\end{figure}


\begin{figure}
\plotone{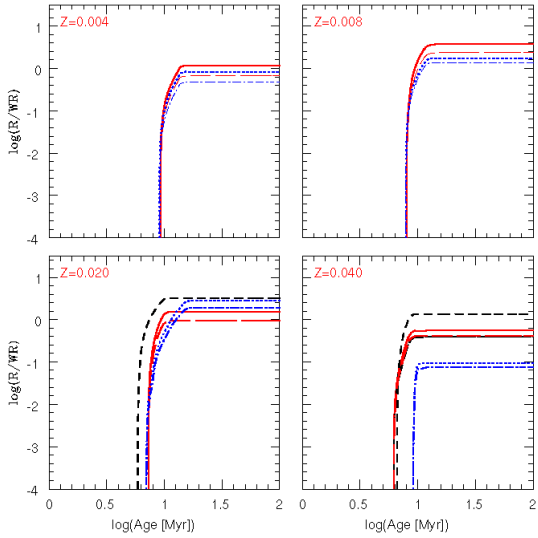}
\caption{Same as Fig.~\ref{fig7}, but for the ratio of RSGs over WR stars.
\label{fig10}}
\end{figure}


\begin{figure}
\plotone{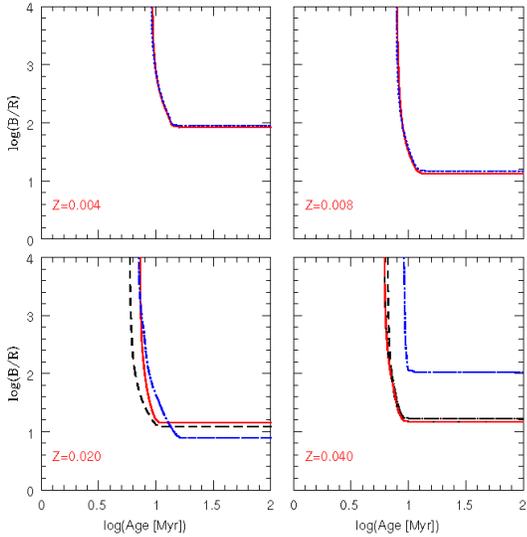}
\caption{Same as Fig.~\ref{fig7}, but for the ratio of blue over red supergiants.
Note that the apparent increase of B/R with decreasing $Z$ seen for $Z=0.004$ and 0.008
is an artifact of incomplete tracks.
\label{fig11}}
\end{figure}


\begin{figure}
\plotone{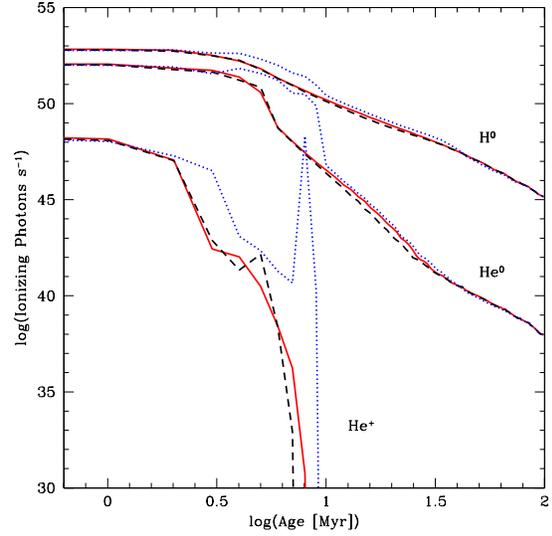}
\caption{Number of photons in the H$^0$, He$^0$, and He$^+$ continuum (solar composition). Line types as in Fig.~\ref{fig3}.
\label{fig12}}
\end{figure}


\begin{figure}
\plotone{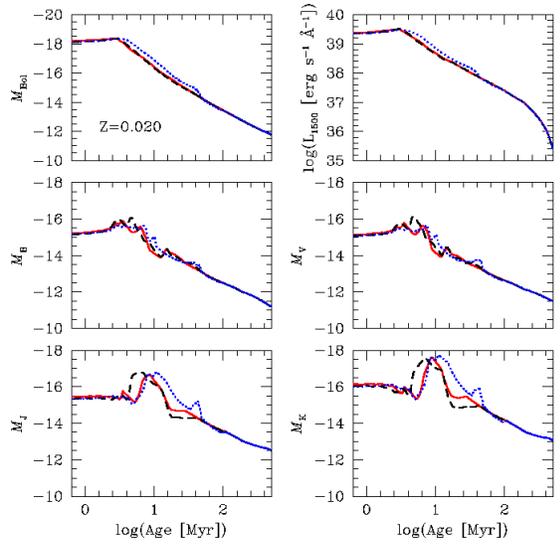}
\caption{$M_{\rm{bol}}$, $L_{\rm{1500}}$, $M_B$, $M_V$, $M_J$, and $M_K$ vs. time for models with the old Geneva (red, solid), new Geneva with {\vrotcero} (black, short-dashed), and new models with {\vrotthree} \kms (blue, dotted) tracks. The bands most affected by the new models are those in the ultraviolet and infrared. Solar chemical composition.
\label{fig13}}
\end{figure}


\begin{figure}
\plotone{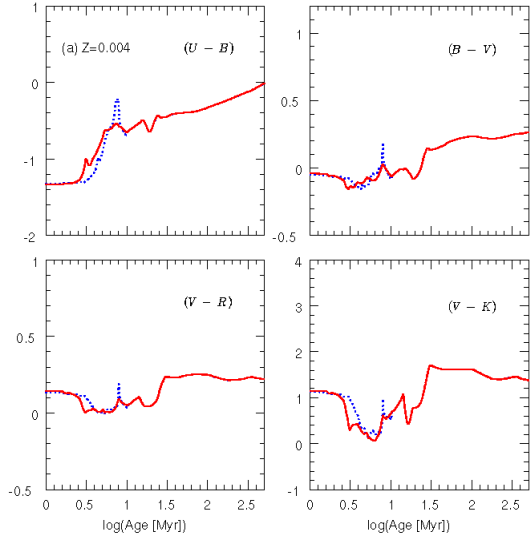}
\caption{Color evolution for selected filters. Lines are as in Fig.~\ref{fig13}. a) $Z = 0.004$; b) $Z = 0.008$; c) $Z = 0.020$. Model with $Z = 0.020$ and the new tracks of {\vrotthree} {\kms} evolve until 100 {\myr} whereas the other models terminate earlier. This is due to the larger mass range covered (9~--~120 {\msun}) than models at lower chemical composition. 
\label{fig14}}
\end{figure}
\begin{figure}
\plotone{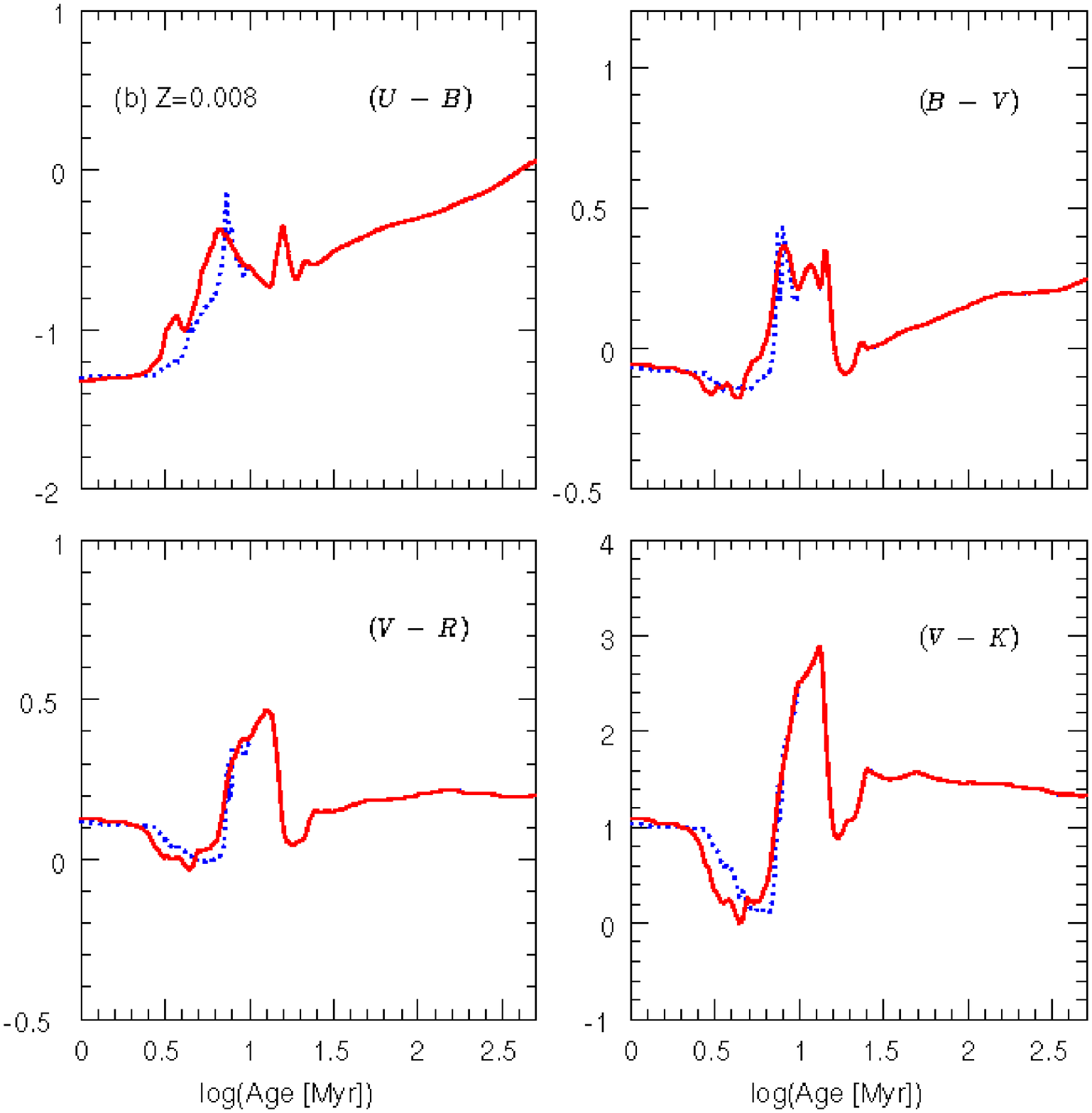}
\end{figure}
\begin{figure}
\plotone{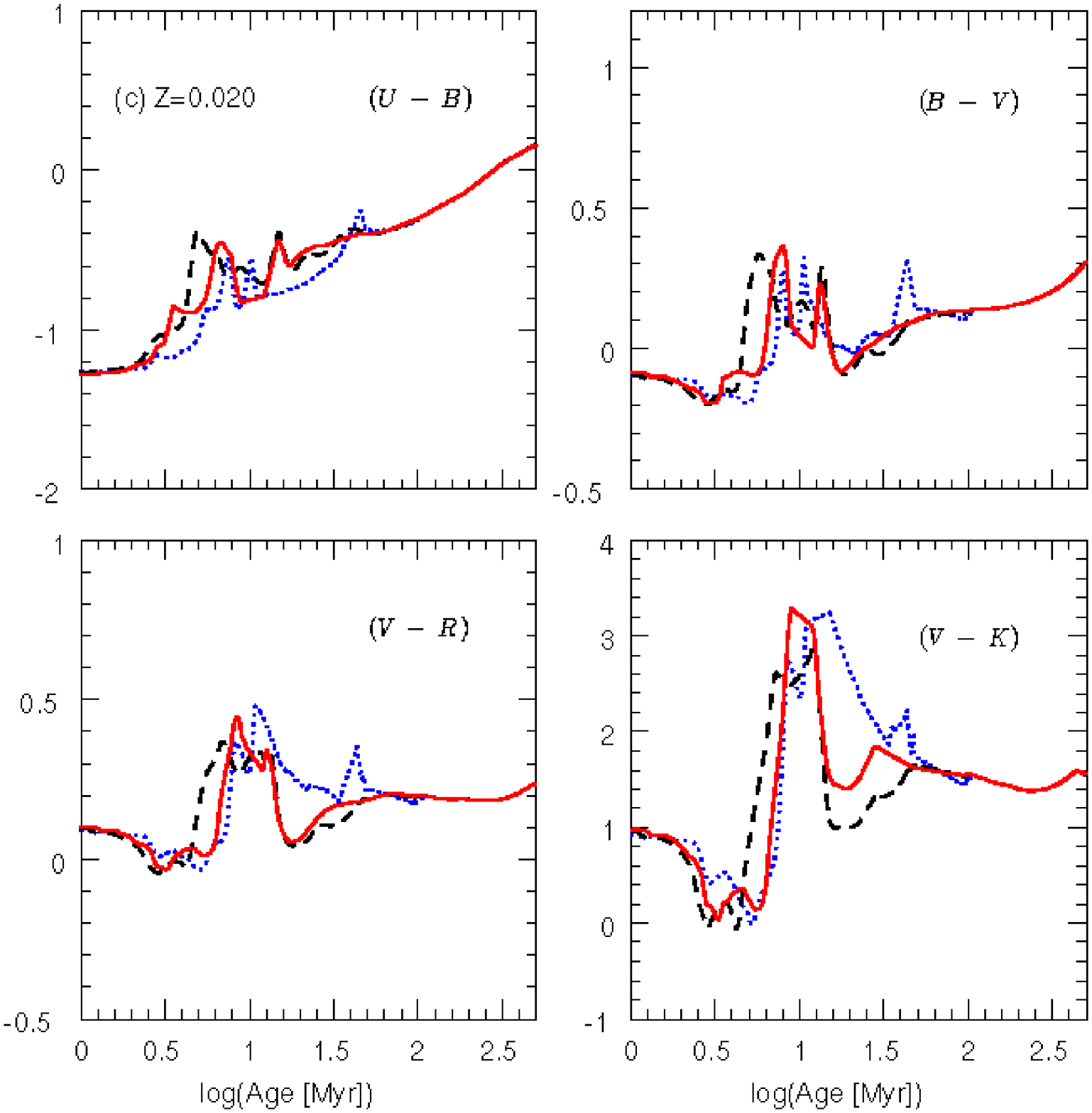}
\end{figure}

\clearpage

\begin{figure}
\plotone{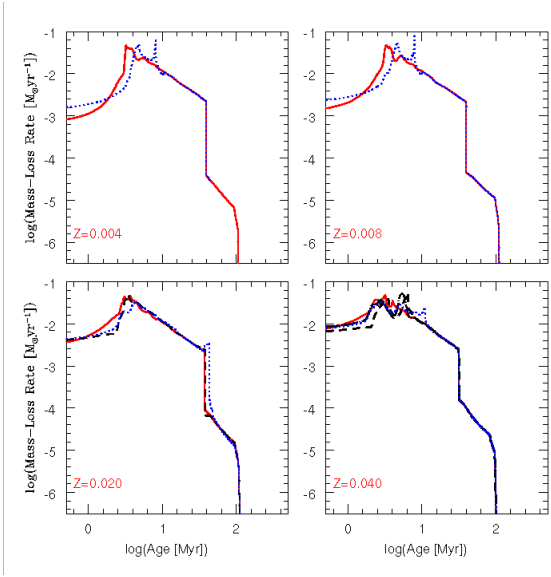}
\caption{Evolution of the mass-loss rate for five sets of models with $10^6$~{\msun}. Four different chemical compositions are plotted. The figure for $Z=0.040$ includes models with a mass-loss rate $\propto Z^{0.5}$ for the WR phase.  Close to the ZAMS, the mass-loss rate is briefly higher in models with {\vrotthree} {\kms} (blue, dotted) than in models with the old tracks (red, solid). The short-dashed lines are models with the new tracks of {\vrotcero} {\kms}. The adopted stellar mass-loss rates for the rotating and non-rotating models are the same. A correction factor has been applied to the rotating models as explained in Sect.~2.1.
\label{fig15}}
\end{figure}


\begin{figure}
\plotone{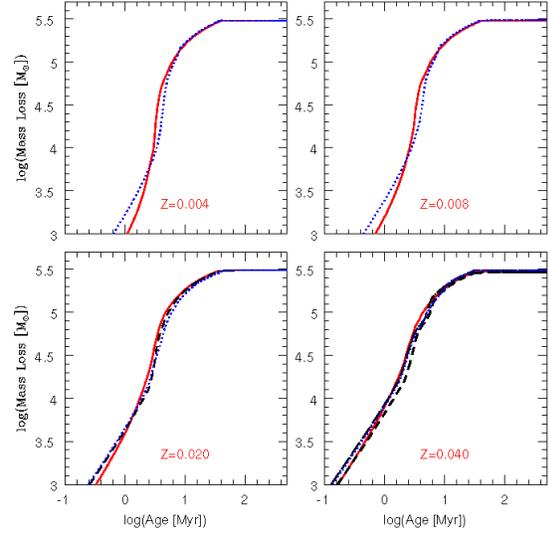}
\caption{Evolution of the mass loss for the same models as in Fig.~\ref{fig15}. The total mass loss is similar in all
models.
\label{fig16}}
\end{figure}


\begin{figure}
\plotone{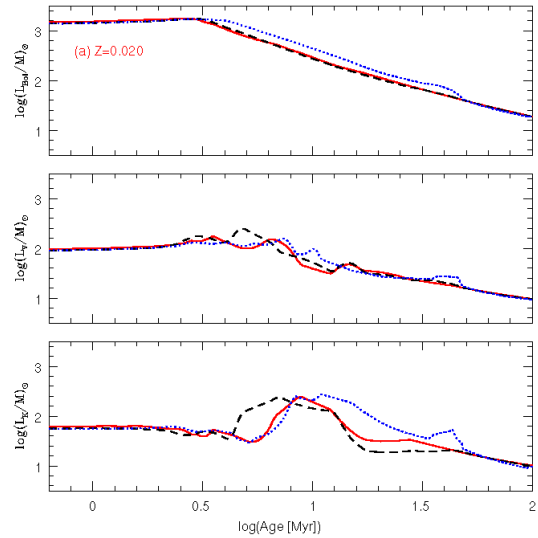}
\caption{The light-to-mass ratio for $L_{\rm{bol}}$, $L_{V}$, and $L_{K}$ for different chemical compositions. a) $Z = 0.020$, and b) $Z = 0.040$. Same line types as before. Models with the new tracks of {\vrotthree} {\kms} are typically more luminous than those with the old Geneva tracks.  
\label{fig17}}
\end{figure}
\begin{figure}
\plotone{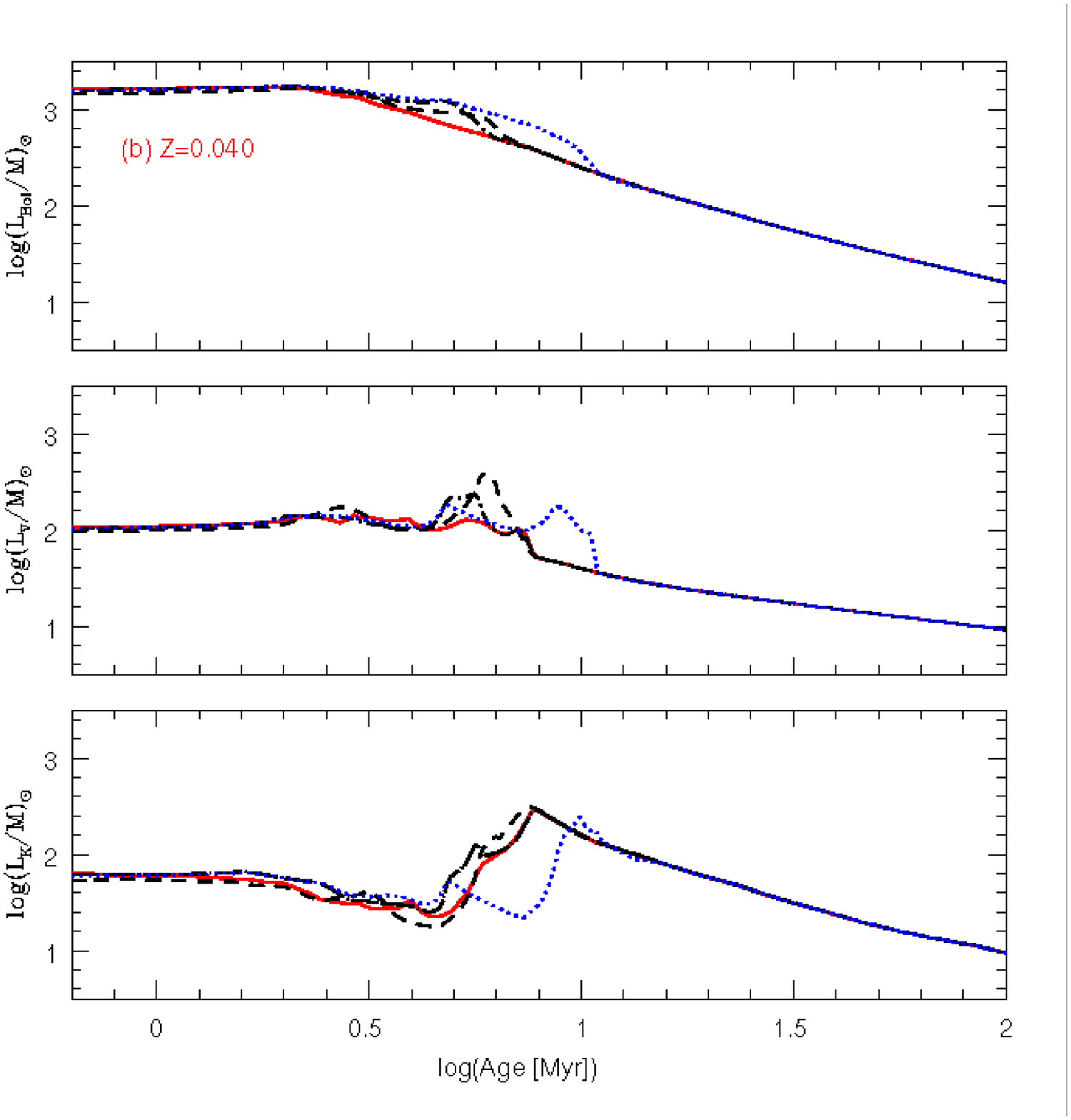}
\end{figure}


\begin{figure}
\plotone{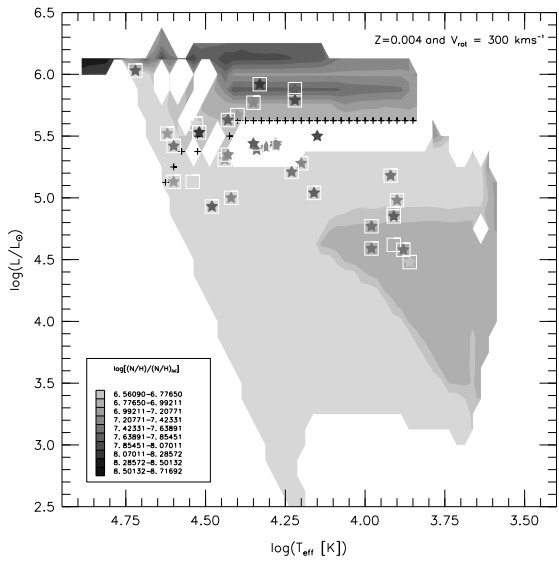}
\caption{Hess-Diagram for $\log[(N/H)/(N/H)_{\rm ini}]$. The crosses indicate the validity limit of the models with the new tracks of {\vrotthree} {\kms}. $Z=0.004$. Symbols denote data. 
\label{fig18}}
\end{figure}


\begin{figure}
\plotone{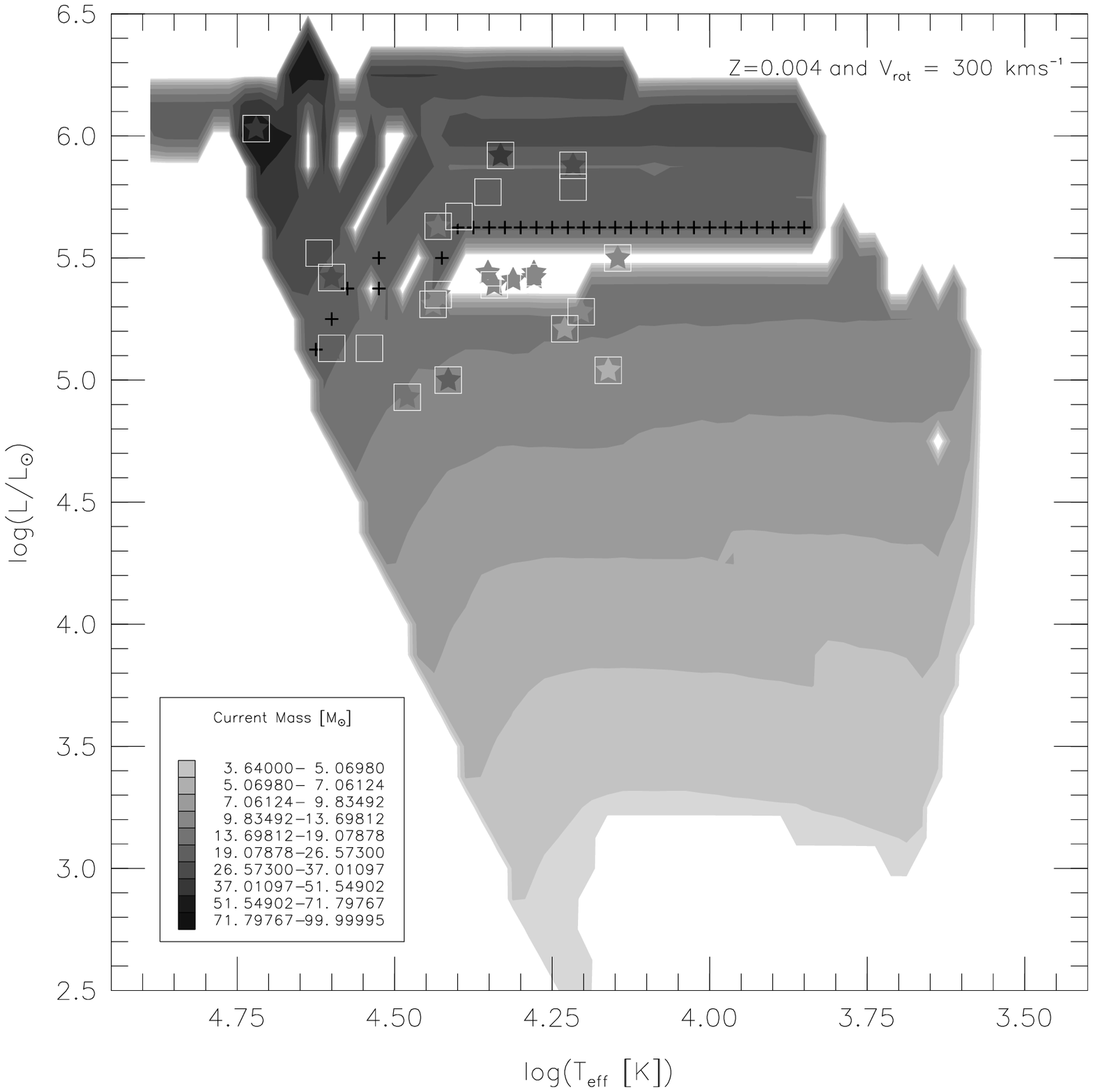}
\caption{Same as in Fig~\ref{fig18}, but for the current mass.
\label{fig19}}
\end{figure}


\begin{figure}
\plotone{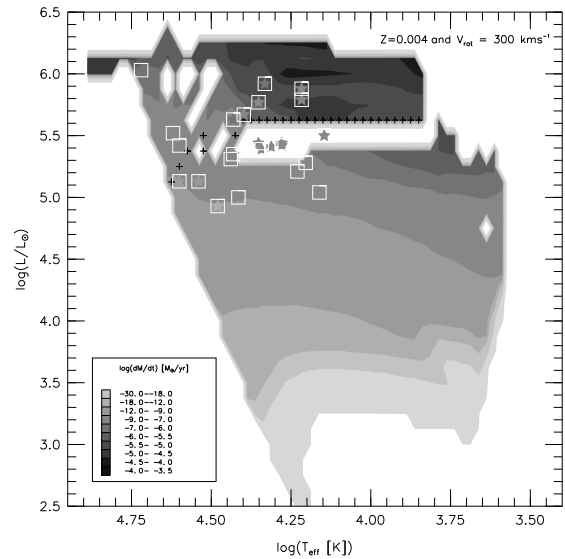}
\caption{Same as in Fig~\ref{fig18}, but for the mass-loss rate.
\label{fig20}}
\end{figure}


\begin{figure}
\plotone{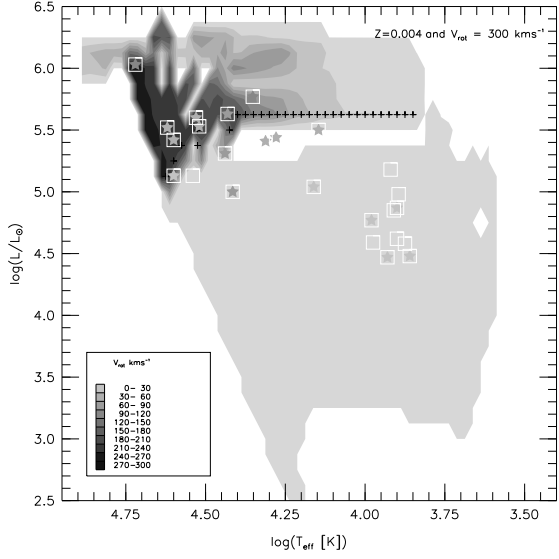}
\caption{Same as in Fig.~\ref{fig18}, but for the rotation velocity.
\label{fig21}}
\end{figure}


\begin{figure}
\plotone{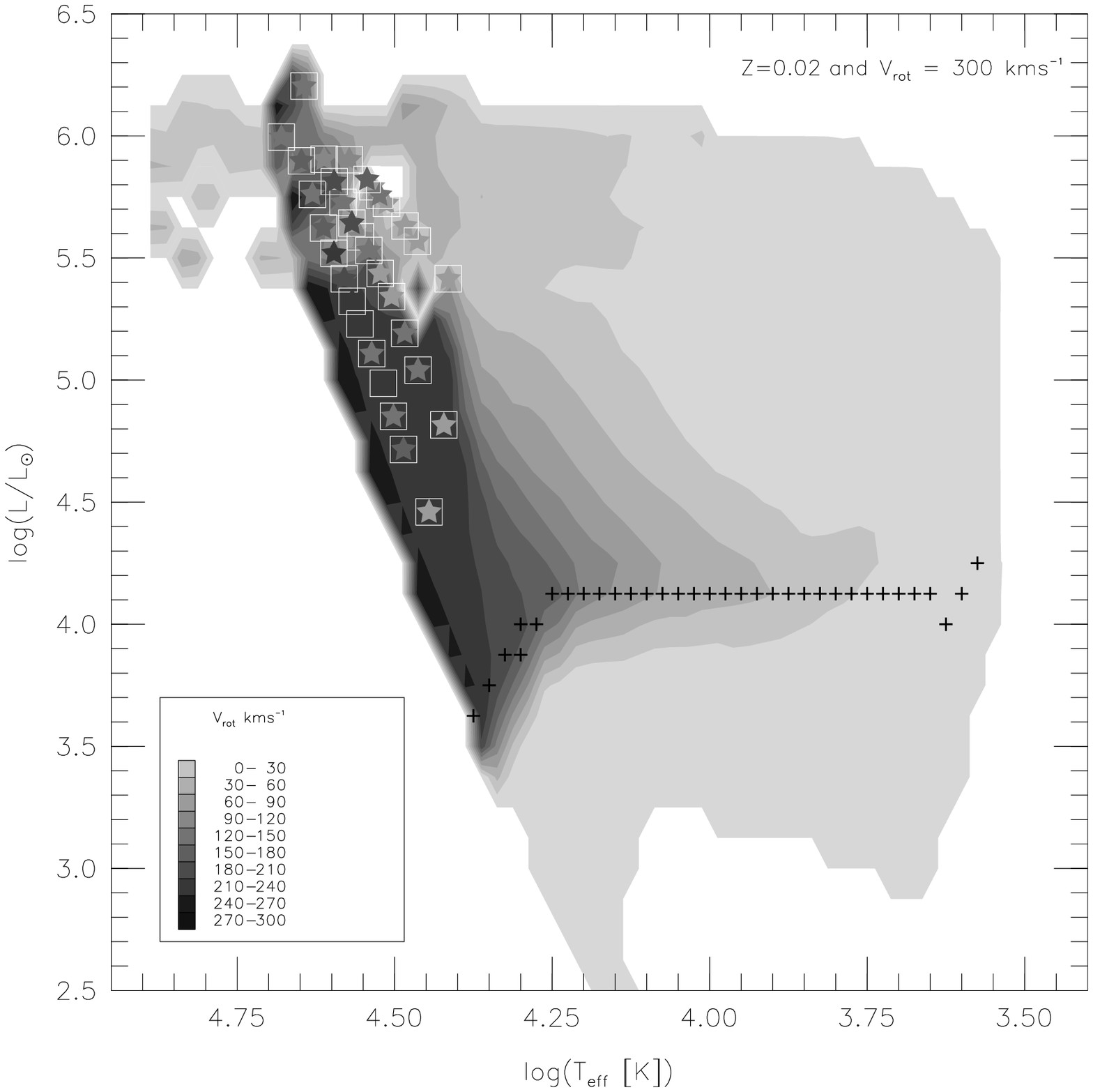}
\caption{Same as in Fig.~\ref{fig21}, but for $Z=0.020$ and data for Galactic O stars.
\label{fig22}}
\end{figure}


\begin{figure}
\plotone{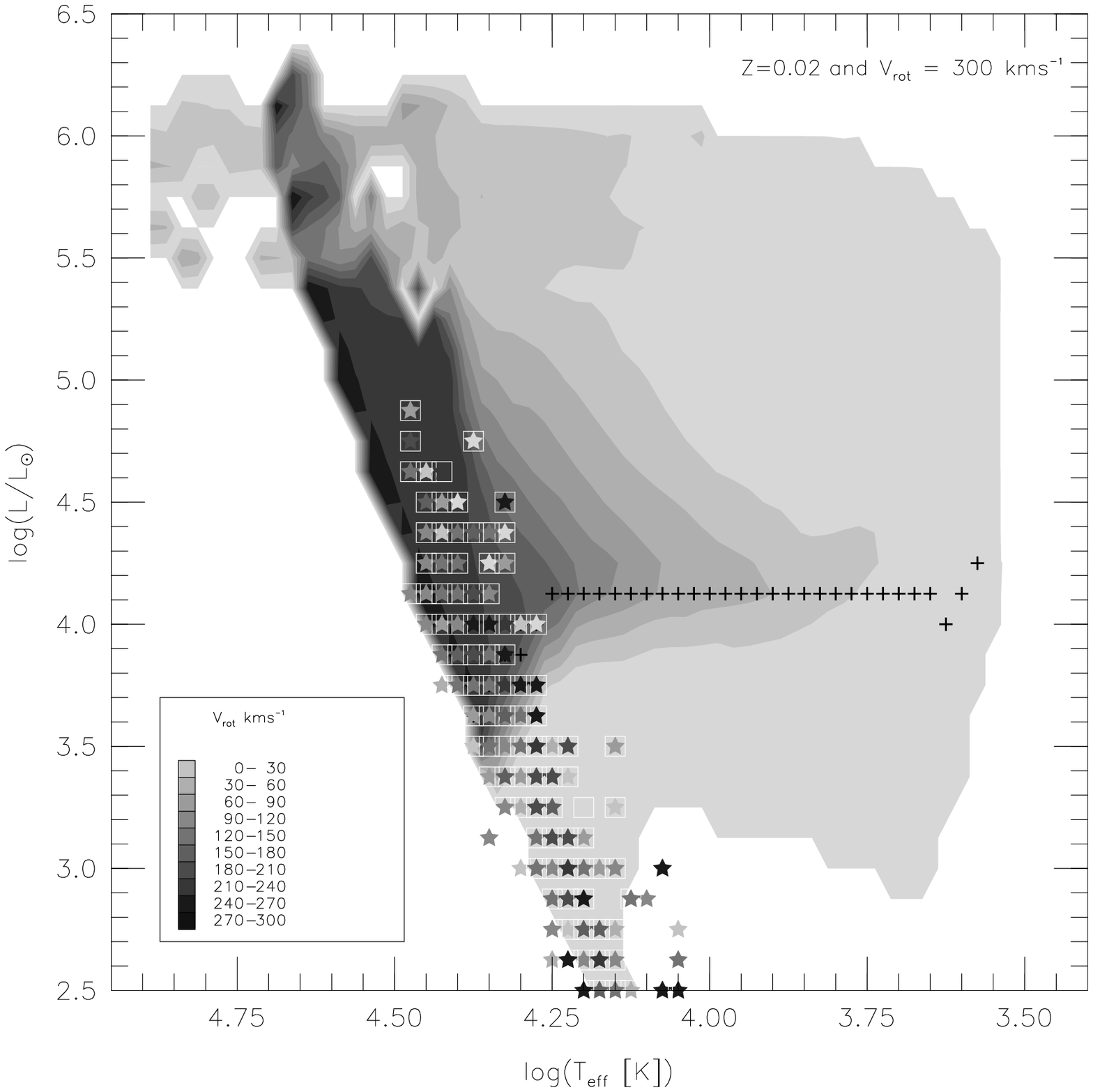}
\caption{Same as in Fig.~\ref{fig21}, but for $Z=0.020$ and data for Galactic B stars.
\label{fig23}}
\end{figure}


\begin{figure}
\plotone{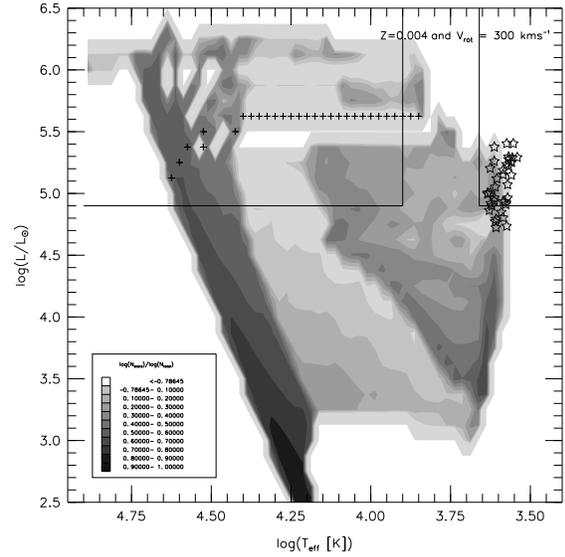}
\caption{Same as in Fig.~\ref{fig24} but with the new RSG sample and $T_{\eff}$ scale of \citet{levesque06}. 
\label{fig24}}
\end{figure}


\begin{figure}
\plotone{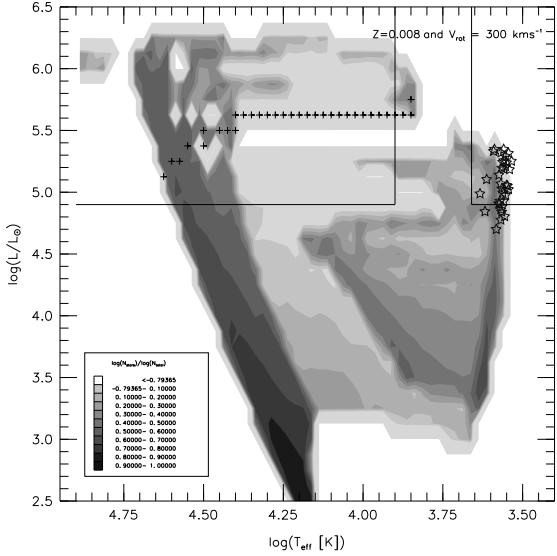}
\caption{Same as in Fig.~\ref{fig24} but for LMC RSGs and models with $Z = 0.008$.
\label{fig25}}
\end{figure}


\begin{figure}
\plotone{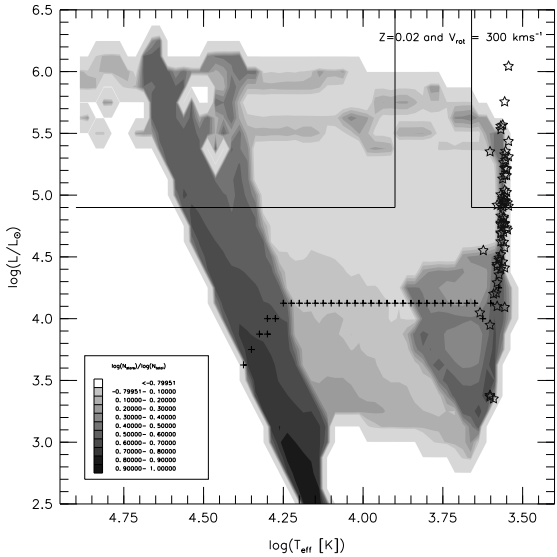}
\caption{Same as in Fig.~\ref{fig25} but for Galactic RSGs and models with $Z = 0.020$. In this case, the $T_{\eff}$ and luminosity limits are from \citet{mas02} and the data from \citet{levesque05}. 
\label{fig26}}
\end{figure}


\begin{figure}
\plotone{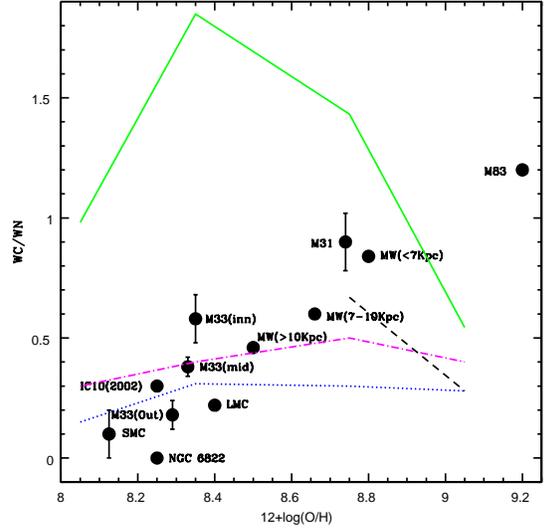}
\caption{Ratio of WC/WN as a function of the oxygen abundance. Models with a temperature limit of $\log T_{\eff} = 4.0$ for entering the WR phase and the old Geneva tracks (solid), new tracks of {\vrotthree} {\kms} (dotted), and new tracks of {\vrotcero} {\kms} (short-dashed) are shown. An additional model with a temperature limit of $\log T_{\eff} = 4.4$ and with the new tracks of {\vrotthree} {\kms} (dot-short dashed) is also shown. The data for the SMC, IC~10, LMC, M~33 and M~31, MW, and M~83 are from \citet{maolpa03,mashol02,breysa99,crockett06,hadfield07} and \citet{hadfield05}, respectively. 
\label{fig27}}
\end{figure}


\begin{figure}
\plotone{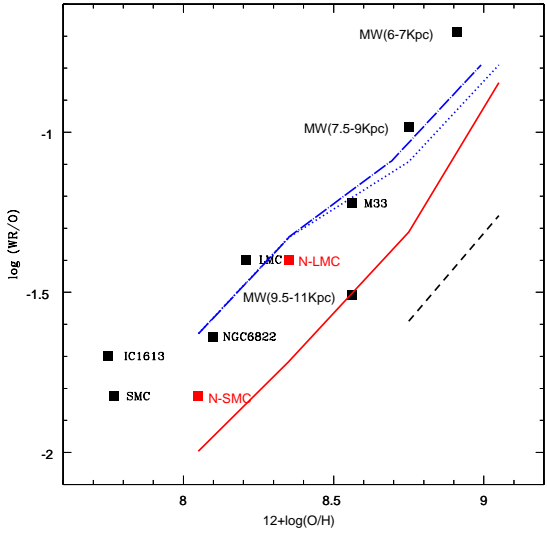}
\caption{Ratio of WR/O as a function of the oxygen abundance. Models are as in Fig.~\ref{fig27} but we have omitted the model with a higher temperature limit for entering the WR phase. The dot-long dashed line shows the trend of the models with the new tracks of {\vrotthree} {\kms} if the oxygen abundance in the models is decreased. The lower oxygen abundance is closer to the observations.
\label{fig28}}
\end{figure}


\begin{figure}
\plotone{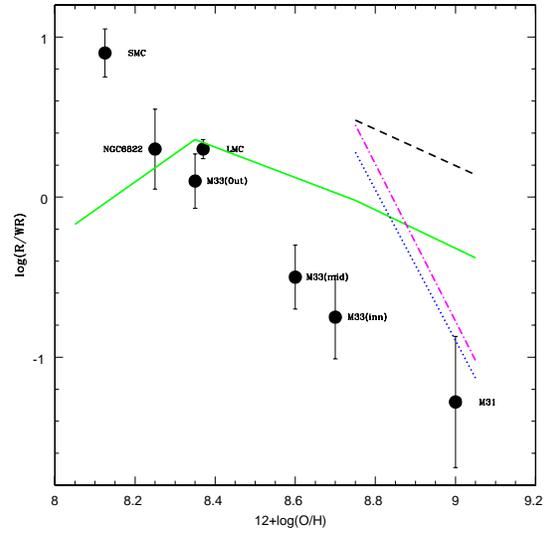}
\caption{Ratio of RSGs over WR stars. Models are as in Fig.~\ref{fig27}. Data are from \citet{mas03}. The new tracks with {\vrotthree} {\kms} for chemical composition lower than solar do not reach RSGs as defined by Massey. For solar and supersolar chemical composition, the models produce too few WR stars to match the observations.
\label{fig29}}
\end{figure}


\begin{figure}
\plotone{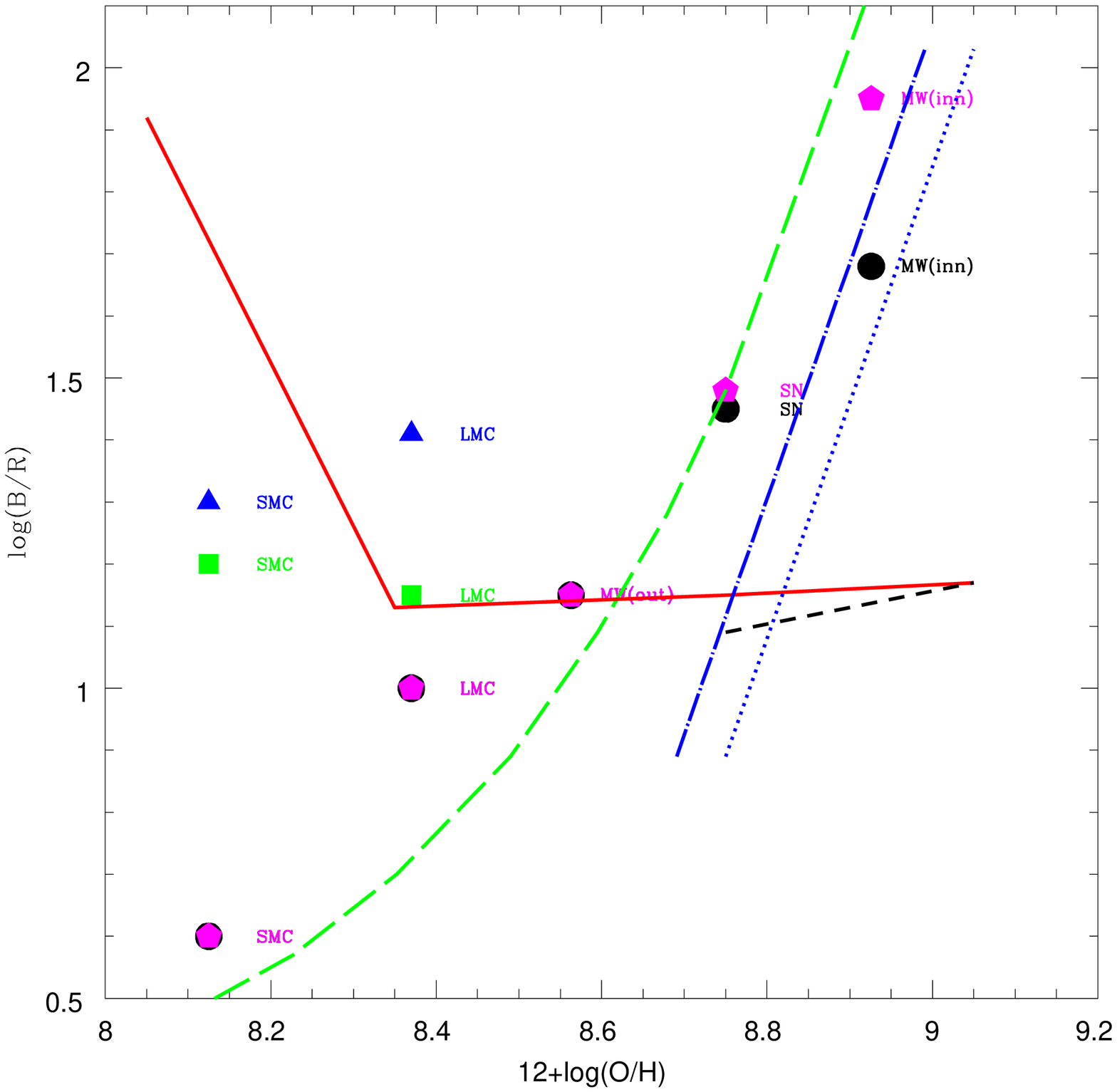}
\caption{Ratio of blue over red supergiants. Models are as in Fig.~\ref{fig28}. Triangles are data from \citet{mas02}, squares are from \citet{maol03}, pentagons are associations of \citet{lanmae95}, circles are field stars of \citet{lanmae95}. The long-dashed line is a fit by \citet{huevo02} to a large sample of clusters in different systems including those from \citet{lanmae95}. The dot-long dashed line denotes models with the new tracks of {\vrotthree} {\kms} using updated values of the solar oxygen abundance. 
\label{fig30}}
\end{figure}

\end{document}